\documentclass[aps,pra,preprint,eqsecnum,groupedaddress,showpacs,floatfix]{revtex4}
\usepackage{dcolumn}
\usepackage{amsmath}
\usepackage{bm}
\usepackage{times}
\newcolumntype{.}{D{x}{}{-1}}

\newcommand*{\centt}[1]{\multicolumn{1}{c}{#1}}
\newcolumntype{w}[1]{D{.}{.}{#1}}

\begin{document}
\preprint{Version 1.0}

\title{$\alpha^4$ Ry corrections to singlet states of helium}

\author{Krzysztof Pachucki}
\email[]{krp@fuw.edu.pl} \homepage[]{www.fuw.edu.pl/~krp}

\affiliation{Institute of Theoretical Physics, Warsaw University,
             Ho\.{z}a 69, 00-681 Warsaw, Poland}

\date{\today}

\begin{abstract}
Corrections of order $\alpha^4$Ry are calculated for the
singlet states $1^1S_0$ and $2^1S_0$ of the helium atom.
The result for  $1^1S_0$ state is in slight
disagreement with that of Korobov and Yelkhovsky in 
[Phys. Rev. Lett. {\bf 87}, 193003 (2001)].
The results obtained lead to a significant improvement of
transition frequencies between low lying levels of the helium atom.
In particular theoretical predictions for the $2^1S_0 - 1^1S_0$ transition 
are found to be in disagreement with experimental values.  
\end{abstract}

 \maketitle

\section{Introduction}
In this paper we present an approach for obtaining precise energy levels of light
few electron atoms and perform calculations for singlet states $1^1S_0$ and
$2^1S_0$ of the helium atom. This approach is based on Quantum Electrodynamic (QED)
theory and relies on expansion in the fine structure constant $\alpha$ \cite{cl, fw}.
It allows one to systematically include nuclear recoil effects, 
electron self-energy, and vacuum polarization. 
Several calculations have already been performed for triplet states of helium, 
the most accurately known being the fine structure of $2^3P_J$, where all corrections up 
to order of $m\,\alpha^7$ have recently been obtained \cite{krp_fs, drake_fs}. 
Other examples include energies of  $1^1S_0$ \cite{ky, ay}, $2^3S_1$
\cite{slamb}, $2^3P$ \cite{plamb}, which have
been obtained up to the order of $m\,\alpha^6$. For other atoms like 
lithium \cite{lithium} or beryllium \cite{beryllium},
energy levels are less accurately known, namely  up to order $m\,\alpha^5$, but are
still
accurate enough to allow the nuclear charge radius to be determined from isotope shift
measurements \cite{lit11}.
 
According to QED theory, the expansion of energy levels in powers of $\alpha$ 
has the following form: 
\begin{equation}
E(\alpha) = E^{(2)} + E^{(4)} + E^{(5)} + E^{(6)} + E^{(7)} + O(\alpha^8),
\end{equation}
where $E^{(n)}$ is a contribution of order $m\,\alpha^n$ and may include 
powers of $\ln\alpha$. Each term $E^{(n)}$ can be expressed as an expectation
value of some effective Hamiltonian or in some cases of a nonlocal operator. 
$E^{(2)}\equiv E_0$ is the eigenvalue of the
nonrelativistic Hamiltonian $H_0$, which for the infinite nuclear mass is
\begin{equation}
H_0 = \sum_a \biggl\{\frac{\vec p_a^{\,2}}{2\,m} -
\frac{Z\,\alpha}{r_a}\biggr\} + \sum_{a>b}\sum_b \frac{\alpha}{r_{ab}}.
\end{equation}
$E^{(4)}$ is the expectation value of the Breit-Pauli Hamiltonian $H^{(4)}$ \cite{bs},
\begin{eqnarray}
H^{(4)} &=&\sum_a \biggl\{-\frac{\vec p^{\,4}_a}{8\,m^3} +
\frac{ \pi\,Z\,\alpha}{2\,m^2}\,\delta^3(r_a)
+\frac{Z\,\alpha}{4\,m^2}\,
\vec\sigma_a\cdot\frac{\vec r_a}{r_a^3}\times \vec p_a\biggr\}
\nonumber \\
&& +\sum_{a>b}\sum_b \biggl\{
-\frac{ \pi\,\alpha}{m^2}\, \delta^3(r_{ab})
-\frac{\alpha}{2\,m^2}\, p_a^i\,
\biggl(\frac{\delta^{ij}}{r_{ab}}+\frac{r^i_{ab}\,r^j_{ab}}{r^3_{ab}}
\biggr)\, p_b^j \nonumber \\ && -
\frac{2\, \pi\,\alpha}{3\,m^2}\,\vec\sigma_a
\cdot\vec\sigma_b\,\delta^3(r_{ab})
+\frac{\alpha}{4\,m^2}\frac{\sigma_a^i\,\sigma_b^j}
{r_{ab}^3}\,
\biggl(\delta^{ij}-3\,\frac{r_{ab}^i\,r_{ab}^j}{r_{ab}^2}\biggr)
+\frac{\alpha}{4\,m^2\,r_{ab}^3} \nonumber \\ &\times& \biggl[
2\,\bigl(\vec\sigma_a\cdot\vec r_{ab}\times\vec p_b -
\vec\sigma_b\cdot\vec r_{ab}\times\vec p_a\bigr)+
\bigl(\vec\sigma_b\cdot\vec r_{ab}\times\vec p_b -
\vec\sigma_a\cdot\vec
r_{ab}\times\vec p_a\bigr)\biggr]\biggr\}\,.
\label{1.3}
\end{eqnarray}
$E^{(5)}$ is the leading QED contribution, which for singlet states is \cite{sap,simple}
\begin{eqnarray}
E^{(5)} &=&\sum_{a>b}\Biggl\langle
\left[\frac{164}{15}+\frac{14}{3}\,\ln\alpha\right]\,
\frac{\alpha^2}{m^2}\,\delta^3(r_{ab})
-\frac{7\,m\,\alpha^5}{6\,\pi}\,{P\left(\frac{1}{(m\,\alpha\,r_{ab})^3}\right)}\Biggr\rangle \\
&& +\sum_a \left[\frac{19}{30}+\ln(\alpha^{-2})-\ln k_0\right]
   \frac{4\,Z\,\alpha^2}{3\,m^2} \langle{\delta^3(r_a)}\rangle, \nonumber
\label{1.4}
\end{eqnarray}
where
\begin{equation} 
\left\langle\phi\left|P\left(\frac{1}{r^3}\right)\right|\psi\right\rangle\equiv
\lim_{a\rightarrow 0}\int d^3 r\, \phi^{*}(\vec r)\,\psi(\vec r)
\left[\frac{1}{r^3}\,\Theta(r-a) + 4\,\pi\,\delta^3(r)\,
(\gamma+\ln a)\right],
\label{1.5}
\end{equation}
with $\Theta$ being the step function and $\gamma$ the Euler constant.
Eq.~(\ref{1.4}) contains the many-electron Bethe logarithm $\ln k_o$ defined by
\begin{equation}
\ln k_0 = \frac{\Bigl\langle\sum_a \vec p_a\,(H_0-E_0)\,
  \ln\bigl[\frac{2\,(H_0-E_0)}{m\,\alpha^2}\bigr]\,
  \sum_b\vec p_b\Bigr\rangle}{2\,\pi\,Z\,\alpha\,
  \Bigl\langle\sum_c\delta^3(r_c)\Bigr\rangle}.
\end{equation}
The calculation of $E^{(6)}$ is the subject of this work. 
It can be represented as
\begin{equation}
E^{(6)} = \bigl\langle H^{(6)} \bigr\rangle + 
\biggl\langle H^{(4)}\,\frac{1}{(E_0-H_0)'}\,H^{(4)}\biggr\rangle,
\end{equation}
where $H^{(6)}$ is the effective Hamiltonian of order $m\,\alpha^6$.
Its derivation is presented in the following section.
Since individual terms in above equation are divergent we follow the 
approach of Ref. \cite{ky} 
and use the technique of dimensional regularization, details of which are presented in 
Appendix A.  $H^{(4)}$ in the above equation is therefore a Breit-Pauli Hamiltonian
in $d$-dimensions, the derivation of which is also included in Section II.
For the next order term $E^{(7)}$ we will use an approximate formula
based on hydrogenic values \cite{eides}.

\section{Derivation of effective Hamiltonian}
To derive the effective Hamiltonian, we follow Ref. \cite{fw}, consider the Dirac equation
with electromagnetic field, and perform a nonrelativistic expansion by the
use of the Foldy-Wouthuysen  transformation, (see Appendix B for details)
\begin{eqnarray}
 H_{\rm FW} &=& e\,A^0 + \frac{\pi^2}{2\,m}-\frac{e}{4\,m}\,\sigma^{ij}\,B^{ij} - 
\frac{\pi^4}{8\,m^3}-\frac{e}{8\,m^2}\Bigl(\vec\nabla\cdot\vec E + 
\sigma^{ij}\,\bigl\{E^i\,,\,\pi^j\bigr\}\Bigr)
\nonumber \\ &&
+\frac{e}{16\,m^3}\bigl\{\sigma^{ij} B^{ij}\,,\,p^2\bigr\}  
-\frac{e}{16\,m^3}\,\bigl\{\vec p\,,\, \partial_t{\vec E}\bigr\}
+\frac{3\,e}{32\,m^4}\,\bigl\{\sigma^{ij}\,E^i\,p^j\,,\,p^2\bigr\}
\nonumber \\ &&
+\frac{e}{128\,m^4}\,[p^2,[p^2,A^0]]
-\frac{3\,e}{64\,m^4}\,\bigl\{p^2\,,\,\nabla^2 A^0 \bigr\} +\frac{p^6}{16\,m^5},
\label{2.1}
\end{eqnarray}
where higher order terms are neglected.
This Hamiltonian defines an effective nonrelativistic QED theory with Lagrangian
\begin{equation}
{\cal L} = \sum_a\psi_a^+(i\,\partial_t-H_{\rm FW})\psi_a + {\cal L}_{EM},
\label{2.2}
\end{equation}
where ${\cal L}_{EM}$ is the Lagrangian of the electromagnetic field and
the summation goes over all particles.
We consider now the equal time retarded Green function $G=G(\{\vec r'_a\}, t'; 
\{\vec r_a\},t)$, where by $ \{\vec r_a\}$ we denote the set of coordinates
for all particles of the system.  In the absence of time dependent perturbation
$G = G(t'-t)$. The Fourier transform of $G$ in the time variable $t'-t$
can be written as
\begin{equation}
G(E) \equiv \frac{1}{E-H_{\rm eff}(E)}, \label{24}
\end{equation}
which is the definition of the effective Hamiltonian $H_{\rm eff}(E)$.
In the nonrelativistic case $H_{\rm eff}=H_0$,
\begin{equation}
H_0 = \frac{\vec p_1^{\,2}}{2\,m} + \frac{\vec p_2^{\,2}}{2\,m}
  -\left[\frac{Z\,\alpha}{r_1}\right]_\epsilon
  -\left[\frac{Z\,\alpha}{r_2}\right]_\epsilon
  + \left[\frac{\alpha}{r_{12}}\right]_\epsilon\,,
\end{equation}
where $[Q]_\epsilon$ is the $d=3-2\,\epsilon$ extension of the operator
$Q$ at $d=3$, see Appendix A for details. All the relativistic and QED
corrections resulting from the Lagrangian $\cal L$ can be represented as
\begin{eqnarray}
G(E) &=& \frac{1}{E-H_0}+\frac{1}{E-H_0}\,\Sigma(E)\,\frac{1}{E-H_0} +
\frac{1}{E-H_0}\,\Sigma(E)\,\frac{1}{E-H_0}\,\Sigma(E)\,\frac{1}{E-H_0}+\ldots
\nonumber \\
&=& \frac{1}{E-H_0-\Sigma(E)} \equiv \frac{1}{E-H_{\rm eff}(E)},\label{25}
\end{eqnarray}
where $\Sigma(E)$ is the $n$-particle irreducible contribution.
The energy level can be interpreted as a pole of $G(E)$
as a function of $E$. It is convenient to 
consider the matrix element of $G$ between
the nonrelativistic wave function corresponding to this 
energy level. There is always such a correspondence, since
relativistic and QED effects are small perturbations of the system. 
This matrix element is
\begin{equation}
\langle\phi|G(E)|\phi\rangle = 
\langle\phi|\frac{1}{E-H_0-\Sigma(E)}|\phi\rangle \equiv 
\frac{1}{E-E_0-\sigma(E)}, \label{26}
\end{equation}
where 
\begin{equation}
\sigma(E) = \langle\phi|\Sigma(E)|\phi\rangle + 
\sum_{n\neq 0}\langle\phi|\Sigma(E)|\phi_n\rangle
\,\frac{1}{E-E_n}\,\langle\phi_n|\Sigma(E)|\phi\rangle
+\ldots
\end{equation}
Having $\sigma(E)$, the correction to the energy level 
can be expressed as
\begin{eqnarray}
\delta E &=& E-E_0 = \sigma(E_0) +\sigma'(E_0)\, \sigma(E_0)+\ldots
\nonumber \\ &=&
\langle\phi|\Sigma(E_0)|\phi\rangle + 
\langle\phi|\Sigma(E_0)\,\frac{1}{(E_0-H_0)'}\,\Sigma(E_0)|\phi\rangle+
\langle\phi|\Sigma'(E_0)|\phi\rangle\,\langle\phi|\Sigma(E_0)|\phi\rangle 
+\ldots \label{2.8}
\end{eqnarray}
Since the last term in Eq. (\ref{2.8}) can be neglected up to order $m\,\alpha^6$, 
one can consider only $\Sigma(E_0)$. In most cases, the explicit state dependence 
of $\Sigma$ can be eliminated by the use of commutation relations.
The only exception is the so called Bethe logarithm, which contributes
only in order $m\,\alpha^5$. If we consider this term separately,
the operator $\Sigma$ gives an effective Hamiltonian
\begin{equation}
H_{\rm eff} = H_0 + \Sigma  = H_0 + H^{(4)} + H^{(5)} +  H^{(6)} + \ldots 
\end{equation}
from which one can calculate corrections to energy levels.
The calculation of $\Sigma$ follows from the Feynman rules for the
Lagrangian in Eq. (\ref{2.2}). 
We will use the photon propagator in the Coulomb gauge:
\begin{eqnarray}
G_{\mu\nu}(k) = \left\{
\begin{array}{ll}
{-\frac{1}{\vec{k}^2}} & \mu = \nu = 0\,, \\
{\frac{-1}{k_0^2-\vec k^2 +i\,\varepsilon}}\Bigl(\delta_{ij}-{\frac{{k}_i {k}_j}
{\vec{k}^2}\Bigr)} & \mu =i, \nu =j
\end{array}
\right.\,, 
\end{eqnarray}
and consider separately corrections due to exchange of the Coulomb $G_{00}$
and the transverse $G_{ij}$ photon. The typical one photon exchange
contribution between electrons $a$ and $b$ is:
\begin{eqnarray}
\langle\phi|\Sigma(E_0)|\phi\rangle &=& e^2\int\frac{d^4
k}{(2\,\pi)^4\,i}\,G_{\mu\nu}(k)\,\biggl\{
\biggl\langle\phi\biggl|\jmath^\mu_a(k)\,e^{i\,\vec k\cdot\vec
r_a} \,\frac{1}{E_0-H_0-k^0+i\,\varepsilon}
\,\jmath^\nu_b(-k)\,e^{-i\,\vec k\cdot\vec
r_b}\,\biggr|\phi\biggr\rangle \nonumber \\ &&
+\biggl\langle\phi\biggl|\jmath^\mu_b(k)\,e^{i\,\vec k\cdot\vec
r_b} \,\frac{1}{E_0-H_0-k^0+i\,\varepsilon}
\,\jmath^\nu_a(-k)\,e^{-i\,\vec k\cdot\vec
r_a}\,\biggr|\phi\biggr\rangle \biggr\}\,,
\label{2.11}
\end{eqnarray}
where $\phi$ is an eigenstate of $H_0$ and $\jmath^\mu_a$ is the
electromagnetic current for particle $a$. 
The first terms of the nonrelativistic expansion of $\jmath^0$ component are
obtained from Eq. (\ref{2.1}) (terms involving coupling to $A^0$)
\begin{equation}
\jmath^0(\vec k) = 1 +\frac{i}{4\,m}\,\sigma^{ij}\,k^i\,p^j 
- \frac{1}{8\,m^2}\vec k^{\,2}+\ldots
\end{equation}
and of the $\vec\jmath$ component (terms involving coupling to $\vec A$)
\begin{equation}
\jmath^i(\vec k) = \frac{p^i}{m} +
\frac{i}{2\,m}\,\sigma^{ji}\,k^j\,.
\label{2.13}
\end{equation}
Most of the calculation is performed in the nonretardation approximation,
namely one sets $k^0=0$ in the photon propagator $G_{\mu\nu}(k)$ and $\jmath(k)$. 
The retardation corrections are considered separately.
Within this approximation and using the symmetrization $k^0 \leftrightarrow -k^0$, 
the $k^0$ integral is
\begin{equation}
\frac{1}{2} \int\frac{d\,k^0}{2\,\pi\,i}\,
\biggl[\frac{1}{-\Delta E-k^0+i\,\varepsilon}+
\frac{1}{-\Delta E+k^0+i\,\varepsilon}\biggr] = -\frac{1}{2},
\end{equation}
which leads to
\begin{equation}
\langle\phi|\Sigma(E_0)|\phi\rangle = -e^2\int\frac{d^d
k}{(2\,\pi)^d}\,G_{\mu\nu}(k_0=0,\vec k)\,
\biggl\langle\phi\biggl|\jmath^\mu_a(\vec k)\,e^{i\,\vec
k\cdot(\vec r_a-\vec r_b)} \,\jmath^\nu_b(-\vec
k)\,\biggr|\phi\biggr\rangle\,.\label{2.15}
\end{equation}
One recognizes that in the nonrelativistic limit $G_{00}$ gives the
Coulomb interaction. However this term is already included
in $H_0$, which means that this nonrelativistic Coulomb interaction
has to be excluded from the perturbative expansion. Next order terms
resulting from $\jmath^0$ and $\vec\jmath$ lead to the Breit
Pauli Hamiltonian $H_{BP}$. This includes corrections to the electric 
as well as magnetic interactions
between electrons and the nucleus. Corrections to the
kinetic energy and electron-nucleus interaction are obtained from (\ref{2.1})
by setting $e\,A^0 = -[(Z\,\alpha)/r_a]_\epsilon$,
\begin{equation}
 \delta_1 H^{(4)} = \sum_{a=1,2} -\frac{p_a^4}{8\,m^3} 
+\frac{\pi\,Z\alpha}{2\,m^2}\,\delta^d(r_a)
-\frac{1}{4\,m^2}\,\sigma_a^{ij}\,\nabla^i
\left[\frac{Z\,\alpha}{r_a}\right]_\epsilon p_a^j.
\label{2.16}
\end{equation}
The derivation of electron-electron interactions is as follows. 
The $\jmath^0$ component gives relativistic corrections of the form
\begin{eqnarray}
\delta_2 H^{(4)} &=& e^2\int\frac{d^d k}{(2\,\pi)^d}\,\frac{1}{\vec k^2}\,
\jmath^0_1(\vec k)\,e^{i\,\vec k\cdot(\vec r_1-\vec r_2)} \,\jmath^0_2(-\vec k) 
\nonumber \\ &=& 
 e^2\int\frac{d^d k}{(2\,\pi)^d}\,\frac{1}{\vec k^2}\,
\biggl(1 + \frac{i}{4\,m}\,\sigma_1^{ij}\,k^i\,p_1^j -\frac{\vec k^2}{8\,m^2}\biggr)
\,e^{i\,\vec k\cdot(\vec r_1-\vec r_2)} \,
\biggl(1 - \frac{i}{4\,m}\,\sigma_2^{ij}\,k^i\,p_2^j -\frac{\vec k^2}{8\,m^2}\biggr)
\nonumber \\ &=& 
- \frac{\pi\,\alpha}{m^2}\,\delta^d(r)
+\frac{1}{4\,m^2}\,\biggl(
 \sigma_1^{ij}\,\nabla^i\left[\frac{\alpha}{r}\right]_\epsilon p_1^j
-\sigma_2^{ij}\,\nabla^i\left[\frac{\alpha}{r}\right]_\epsilon p_2^j
\biggr)\,.
\label{2.17}
\end{eqnarray}
where $ r \equiv r_{12} = |\vec r_1-\vec r_2|$. We have left out in the above
pure Coulomb interaction between electrons and neglected higher order terms.
The $\vec\jmath$ component gives the following corrections:
\begin{eqnarray}
\delta_3 H^{(4)} &=& -e^2\int\frac{d^d k}{(2\,\pi)^d}\,\frac{1}{\vec k^2}\,
\left(\delta^{ij}-\frac{k^i\,k^j}{\vec k^2}\right)
\jmath^i_1(\vec k)\,e^{i\,\vec k\cdot(\vec r_1-\vec r_2)} \,\jmath^j_2(-\vec k) 
\nonumber \\ &=& 
-e^2\int\frac{d^d k}{(2\,\pi)^d}\,\frac{1}{\vec k^2}\,
\left(\delta^{ij}-\frac{k^i\,k^j}{\vec k^2}\right)
\left(\frac{p_1^i}{m}+\frac{i}{2\,m}\,\sigma_1^{ij}\,k^j\right)
\,e^{i\,\vec k\cdot(\vec r_1-\vec r_2)}\,
\left(\frac{p_2^i}{m}-\frac{i}{2\,m}\,\sigma_2^{ij}\,k^j\right)
\nonumber \\ &=& 
 -\frac{\alpha}{2\,m^2}\,p_1^i\,
\biggl[\frac{\delta^{ij}}{r}+\frac{r^i\,r^j}{r^3}
\biggr]_\epsilon\,p_2^j
+\frac{1}{4\,m^2}\,\sigma_1^{ik}\,\sigma_2^{jk}\,
\left(\nabla^i\,\nabla^j-\frac{\delta^{ij}}{d}\,\nabla^2\right)\,
\left[\frac{\alpha}{r}\right]_\epsilon
\nonumber \\ &&
- \frac{\pi\,\alpha}{d\,m^2}\,\sigma_1^{ij}\,\sigma_2^{ij}\,\delta^d(r) 
-\frac{1}{2\,m^2}\,\biggl(\sigma_1^{ij}\,\nabla^i\left[\frac{\alpha}{r}\right]_\epsilon\,p^j_2
-\sigma_2^{ij}\,\nabla^i\left[\frac{\alpha}{r}\right]_\epsilon\,p^j_1 \biggr)\,.
\label{2.18}
\end{eqnarray}
The complete relativistic correction $H^{(4)}$ is a sum of 
Eqs. (\ref{2.16}), (\ref{2.17}), (\ref{2.18}),
\begin{equation}
H^{(4)} = \delta_1 H^{(4)} +  \delta_2 H^{(4)} +  \delta_3 H^{(4)}.
\end{equation}
Since spin-orbit terms do not lead to divergences in the second order matrix element
one can assume for them $\epsilon=0$.
The spin-spin tensor interaction, the 2nd term of Eq. (\ref{2.18}), 
vanishes for singlet states, as the total spin is zero, 
see Eq. (\ref{3.29}) for the definition of a singlet state in $d$-dimensions.
Moreover, as a result of this definition one obtains 
$\sigma_1^{ij}\,\sigma_2^{ij} \rightarrow -d(d-1)$, so $H^{(4)}$ becomes
\begin{eqnarray}
H^{(4)} &=& H_A + H_C, \\
H_A &=& -\frac{p_1^4}{8\,m^3} - \frac{p_2^4}{8\,m^3} 
+\frac{\pi\,Z\alpha}{2\,m^2}\,\delta^d(r_1)
+\frac{\pi\,Z\alpha}{2\,m^2}\,\delta^d(r_2)\nonumber \\ &&
-\frac{\alpha}{2\,m^2}\,p_1^i\,
\biggl[\frac{\delta^{ij}}{r}+\frac{r^i\,r^j}{r^3}
\biggr]_\epsilon\,p_2^j
+\frac{\pi\,\alpha}{m^2}\,(d-2)\,\delta^d(r),
\\
H_C &=& \frac{(\vec\sigma_1-\vec\sigma_2)}{2}\,
\biggl[\frac{Z}{4\,m^2}\biggl(\frac{\vec r_1}{r_1^3}\times\vec p_1-\frac{\vec
    r_2}{r_2^3}\times\vec p_2\biggr)
+\frac{1}{4\,m^2}\,\frac{\vec r}{r^3}\,\times(\vec p_1+\vec p_2)\biggr],
\end{eqnarray}
in agreement with Ref. \cite{ay}.
Both $H_A$ and $H_C$ contribute to $E^{(6)}$ through second order
contribution, namely
\begin{eqnarray}
E_A &=& \biggl\langle H_A\,\frac{1}{(E_0-H_0)'}\,H_A\biggr\rangle, \\
E_C &=& \biggl\langle H_C\,\frac{1}{E_0-H_0}\,H_C\biggr\rangle, \label{2.24}
\end{eqnarray}

Below we derive the higher order
term in the nonrelativistic expansion, namely
the $m\,\alpha^6$ Hamiltonian, which we call here the
higher order effective Hamiltonian $H^{(6)}$. 
It is expressed as a sum of various contributions 
\begin{equation}
H^{(6)} =\sum_{i=0,8} \delta H_i + H_H + H_R,
\end{equation}
which are calculated in the following on the basis of the Foldy-Wouthuysen
from Eq. (\ref{2.1}). Similar derivation has already been performed
for the case $d=3$ in Ref. \cite{fw}.
One can neglect here all spin-orbit terms and tensor spin-spin
interactions, as they vanish for singlet states.
$\delta H_{0}$ is the  kinetic energy correction, the last term in Eq. (\ref{2.1}),
\begin{equation}
\delta H_{0} = \frac{p_1^6}{16\,m^5} + \frac{p_2^6}{16\,m^5}\,.
\end{equation}

$\delta H_{1}$ is a correction due to the 9$^{\rm th}$ and 10$^{\rm th}$ terms
in $H_{\rm FW}$ in Eq. (\ref{2.1}). These terms involve only $A^0$, 
so the nonretardation approximation is strictly valid here. 
This correction $\delta H_{1}$ includes the Coulomb interaction
between the electron and the nucleus, and between electrons.
So, if we denote by  $V$ the nonrelativistic interaction potential 
\begin{equation}
V \equiv  \left[-\frac{Z\,\alpha}{r_1}
                -\frac{Z\,\alpha}{r_2}
                +\frac{\alpha}{r}\right]_\epsilon,
\end{equation}
and for later use, by ${\cal E}_a$ the static electric field at the position 
of particle $a$ which is produced by the nucleus and the other particle $b$
\begin{equation}
e\,\vec{\cal E}_a \equiv -\nabla_a V \equiv 
\left[-Z\,\alpha\,\frac{\vec r_a}{r_a^3} + \alpha\,\frac{\vec r_{ab}}{r_{ab}^3}\right]_\epsilon,
\end{equation}
then $\delta H_{1}$ can be written as
\begin{equation}
\delta H_{1} = \sum_{a=1,2}
\frac{1}{128\,m^4}\,[p_a^2,[p_a^2,V]]
-\frac{3}{64\,m^4}\,\Bigl\{p_a^2\,,\,\nabla_a^2 V\Bigr\}.
\end{equation}

$\delta H_{2}$ is a correction 
to the Coulomb interaction between electrons
which comes from the 5$^{\rm th}$ term in $H_{\rm FW}$ Eq. (\ref{2.1}), namely
\begin{equation}
-\frac{e}{8\,m^2}\Bigl(\vec\nabla\cdot\vec E + \sigma^{ij}\,\bigl\{E^i\,,\,p^j\bigr\}\Bigr).
\end{equation}
If interaction of both electrons is modified by this term,
it can be obtained in the nonretardation approximation Eq. (\ref{2.15}), namely
\begin{eqnarray}
\delta H_{2} &=& 
\int \frac{d^d k}{(2\,\pi)^d}\,\frac{4\,\pi}{k^2}\,\frac{1}{64\,m^4}\,
\biggl(k^2 -2\,i\,\sigma_1^{ij}\,k^i_1\,p_1^j\biggr)\,
e^{i\,\vec k\cdot\vec r}\,
\biggl(k^2 +2\,i\,\sigma_2^{kl}\,k^k\,p_2^l\biggr)
\nonumber \\ &=& 
\frac{1}{64\,m^4}\,\biggl\{-4\,\pi\,\alpha\,\nabla^2\,\delta^d(r) +
\frac{16\,\pi\,\alpha}{d(d-1)}\,\sigma_1\,\sigma_2\;
p^i_1\,\bigl[\delta^{ij}_\perp (r)\bigr]_\epsilon\,p_2^j\biggr\},
\end{eqnarray}
where $[\delta^{ij}_\perp (r)]_\epsilon$ is defined in Eq. (\ref{A30})
and we use the identity which is valid for singlet states
\begin{eqnarray}
\sigma_1^{ij}\,\sigma_2^{kl} &=& \sigma_1\,\sigma_2\,
\frac{(\delta^{ik}\,\delta^{jl}-\delta^{il}\,\delta^{jk})}{d(d-1)},\\
\sigma_1\,\sigma_2 &\equiv& \sigma_1^{ij}\,\sigma_2^{ij}.
\end{eqnarray}

$\delta H_{3}$ is the correction that comes from 7$^{\rm th}$ term in Eq. (\ref{2.1}) 
\begin{equation}
-\frac{e}{16\,m^3}\,\bigl\{\vec p\,,\, \partial_t{\vec E}\bigr\}.
\end{equation}
To calculate this correction, we have to return 
to the original expression for one-photon exchange Eq. (\ref{2.11}). We assume
that particle $1$ interacts with electromagnetic field by this term, 
while particle $2$ by nonrelativistic coupling $e\,A^0$ and obtain
\begin{eqnarray}   
\delta E_{3} &=& -e^2\int\frac{d^Dk}{(2\,\pi)^D\,i}\,
\frac{1}{\vec k^2}\,\frac{1}{16\,m^3}\,
\biggl(\langle\phi|\bigl\{\vec p_a\,,\,\vec k\,e^{i\,\vec k\cdot \vec r_a}\bigr\}\,
\frac{k^0}{E_0-H_0-k^0+i\,\varepsilon}\,e^{-i\,\vec k\cdot \vec r_b}|\phi\rangle
\nonumber \\ &&
-\langle\phi|e^{-i\,\vec k\cdot \vec r_b}\,
\frac{k^0}{E_0-H_0-k^0+i\,\varepsilon}\,
\bigl\{\vec p_a\,,\,\vec k\,e^{i\,\vec k\cdot \vec r_a}\bigr\}\,
|\phi\rangle\biggr) + (1\leftrightarrow 2).
\end{eqnarray}
After performing the $k^0$ integral, and commuting $(H_0-E_0)$ with $e^{- i\,\vec k\cdot \vec r_b}$
one expresses this correction in terms of an effective operator
\begin{equation}
\delta H_3 = -\frac{1}{16\,m^4}\,
\left[p_2^2,\left[p_1^2,\left[\frac{\alpha}{r}\right]_\epsilon\right]\right].
\end{equation}

$\delta H_{4}$ is the relativistic correction 
to transverse photon exchange. 
The first electron is coupled to $\vec A$ by the nonrelativistic term
\begin{equation}
-\frac{e}{m}\,\vec p\cdot\vec A -\frac{e}{4\,m}\,\sigma^{ij}\,B^{ij},
\end{equation} 
and the second one by the relativistic correction, 
the 4$^{\rm th}$ and 6$^{\rm th}$ terms in Eq. (\ref{2.1})
\begin{equation}
-\frac{1}{8\,m^3}\,\bigl(\pi^4-\frac{e}{2}\,\bigl\{\sigma^{ij}\,B^{ij}\,,\,p^2\bigr\}\bigr)
\rightarrow
\frac{e}{8\,m^3}\,\bigl\{p^2\,,\,2\,\vec p\cdot\vec A
+\frac{1}{2}\,\sigma^{ij}\,B^{ij}\bigr\}.
\end{equation}
It is sufficient to calculate it in the nonretardation approximation
\begin{eqnarray}
\delta H_4 &=& \frac{\alpha}{8\,m^3}\,
\Bigl[2\,p_1^2\,p_1^i + p_1^2\,\sigma_1^{li}\,\nabla_1^l\Bigr]
\Bigl[\frac{p_2^j}{m} + \frac{1}{2\,m}\,\sigma_2^{kj}\times\,\nabla_2)^j\Bigr]
\nonumber \\ &&
\times\biggl[\frac{1}{2\,r}\biggl(\delta^{ij}+\frac{r^i\,r^j}{r^2}\biggr)\biggr]_\epsilon
+ {\rm h.c.} + (1 \leftrightarrow 2).
\end{eqnarray}
It is convenient at this point to introduce a notation
for the vector potential at the position of particle  $a$ 
which is produced by particle $b$
\begin{equation}
e\,{\cal A}^i_{a} \equiv  \biggl[\frac{\alpha}{2\,r_{ab}}
\biggl(\delta^{ij}+\frac{r_{ab}^i\,r_{ab}^j}{r_{ab}^2}\biggr)\,\frac{p_b^j}{m} +
\frac{\alpha}{2\,m}\,\sigma_b^{ki}\,\frac{r_{ab}^k}{r_{ab}^3}\biggr]_\epsilon\,.
\end{equation}
This correction can then be written as
\begin{eqnarray}
\delta H_4 &=&
\sum_{a=1,2}\,\frac{e}{8\,m^3}\,
\Bigl\{p_a^2\,,\,2\,\vec p_a\cdot\vec {\cal A}_{a} 
+ \sigma_a^{ij}\nabla_a^i\,{\cal A}_{a}^j\Bigr\}
\nonumber \\ &=&
\frac{(p_1^2+p_2^2)}{2}\,p_1^i\,\biggl[\frac{\alpha}{2\,r}\,
\biggl(\delta^{ij}+\frac{r^i\,r^j}{r^2}\biggr)\biggr]_\epsilon\,p_2^j+
\frac{(p_1^2+p_2^2)}{8}\,\frac{\sigma_1\,\sigma_2}{d}\,4\,\pi\,\delta^d(r).
\end{eqnarray}
One notices that in the nonretardation approximation  any correction
can be simply obtained by replacing the magnetic field $\vec A$ by 
a static field $\vec {\cal A}_{a}$. We will use this fact in further calculations.

$\delta H_{5}$ comes from the coupling
\begin{equation}
\frac{e^2}{4\,m^2}\,\sigma^{ij}\,E^i\,A^j,
\end{equation}
 which is present in the 5$^{\rm th}$ term in Eq. (\ref{2.1}). 
The resulting correction is obtained by replacing the fields $\vec E$ and
$\vec A$ by the static fields produced by the other electron
\begin{eqnarray}
\delta H_5 &=& \sum_a \frac{e^2}{4\,m^2}\,\sigma_a^{ij}
{\cal E}_a^i\,{\cal A}_{a}^j =
-\frac{Z\,\alpha^2}{8\,m^3}\frac{\sigma_1\,\sigma_2}{d}\,
\biggl[\frac{\vec r_1}{r_1^3}-\frac{\vec r_2}{r_2^3}\biggr]_\epsilon\,
\biggl[\frac{\vec r}{r^3}\biggr]_\epsilon 
+\frac{\alpha^2}{4\,m^3}\,\frac{\sigma_1\,\sigma_2}{d}\,\biggl[\frac{1}{r^4}\biggr]_\epsilon,
\end{eqnarray}
where $[1/r^4]_\epsilon \equiv (\nabla[1/r]_\epsilon)^2$.

$\delta H_{6}$ comes from the coupling
\begin{equation}
\frac{e^2}{2\,m}\,\vec A^2,
\end{equation}
which is present in 2$^{\rm nd}$ term of Eq. (\ref{2.1}). Again, in the nonretardation
approximation the $\vec A_a$ field is being replaced by the static fields produced
by the other electron
\begin{eqnarray}
\delta H_6 &=& \sum_a\frac{e^2}{2\,m^2}\,\vec {\cal A}_a^2 \nonumber \\ &=&
\frac{\alpha^2}{8}\,p_1^i\,\frac{1}{r^2}\,\biggl(\delta^{ij}+3\frac{r^i\,r^j}{r^2}\biggr)\,p_1^j
+\frac{\alpha^2}{8}\,p_1^i\,\frac{1}{r^2}\,\biggl(\delta^{ij}+3\frac{r^i\,r^j}{r^2}\biggr)\,p_1^j
+\frac{d-1}{4}\,\biggl[\frac{\alpha^2}{r^4}\biggr]_\epsilon,
\end{eqnarray}
where one used the identity
\begin{equation}
\sigma^{ij}\,\sigma^{ij} = d\,(d-1).
\end{equation}

$\delta H_{7}$ is a retardation correction in the nonrelativistic single
transverse photon exchange. To calculate this correction, we have to again return to the 
general one-photon exchange expression, Eq. (\ref{2.11}), and take the transverse
part of the photon propagator
\begin{eqnarray}
\delta E &=& -e^2\,\int\frac{d^Dk}{(2\,\pi)^D\,i}\,
\frac{1}{(k^0)^2-\vec k^2+i\,\varepsilon}\,\biggl(\delta^{ij}-\frac{k^i\,k^j}
{\vec k^2}\biggr)\,
\nonumber \\ &&
\biggl\langle\phi\biggl|\jmath^i_1(k)\,e^{i\,\vec k\cdot\vec
r_1} \,\frac{1}{E_0-H_0-k^0+i\,\varepsilon}
\,\jmath^j_2(-k)\,e^{-i\,\vec k\cdot\vec
r_2}\,\biggr|\phi\biggr\rangle +(1\leftrightarrow 2)\,.
\end{eqnarray}
We assume that the product $ \jmath^i_1(k)\;
\jmath^j_2(-k)$ contains at most a single power of $k^0$.
This allows one to perform the $k^0$ integration by encircling  
the only pole $k^0 = |\vec k|$ on $\Re(k^0)>0$ complex half plane
and obtain
\begin{eqnarray}
\delta E &=& e^2\,\int\frac{d^dk}{(2\,\pi)^d\,2\,k}\,
\biggl(\delta^{ij}-\frac{k^i\,k^j}{k^2}\biggr)\,
\nonumber \\ &&
\biggl\langle\phi\biggl|\jmath^i_1(\vec k)\,e^{i\,\vec k\cdot\vec r_1} \,
\frac{1}{E_0-H_0-k}
\,\jmath^j_2(-\vec k)\,e^{-i\,\vec k\cdot\vec
r_2}\,\biggr|\phi\biggr\rangle +(1\leftrightarrow 2)\,,
\label{2.48}
\end{eqnarray}
where $k = |\vec k|$. By using the nonrelativistic
form of $\jmath^i$ and performing the retardation expansion, 
\begin{equation} 
\frac{1}{E_0-H_0-k} = -\frac{1}{k}+\frac{H_0-E_0}{k^2}
-\frac{(H_0-E_0)^2}{k^3}+\ldots\label{2.49}
\end{equation}
where the first one contributes to the Breit-Pauli Hamiltonian, 
the second term to $E^{(5)}$, and the third term gives $\delta E_{7}$
\begin{eqnarray}
\delta E_7 &=& -e^2\,\int\frac{d^dk}{(2\,\pi)^d\,2\,k^4}\,
\biggl(\delta^{ij}-\frac{k^i\,k^j}{k^2}\biggr)\,
\biggl\langle\phi\biggl|
\biggl(\frac{ p_1^i}{m}+\frac{1}{2\,m}\,
\sigma_1^{ki}\,\nabla_1^k\biggr)
\,e^{i\,\vec k\cdot\vec r_1}\,
\nonumber \\ &&
(H_0-E_0)^2
\,\biggl(\frac{p_2^j}{m}+\frac{1}{2\,m}\,
\sigma_2^{lj}\,\nabla_2^l\biggr)
\,e^{-i\,\vec k\cdot\vec
r_2}\,\biggr|\phi\biggr\rangle + (1\leftrightarrow 2)\,.
\end{eqnarray}
This is the most complicated term in the evaluation.
After $k$-integration one obtains
\begin{eqnarray}
\delta H_7 &=& \frac{\alpha}{16\,m^4}\,\frac{\sigma_1\,\sigma_2}{d}\,
\biggl[p_1^2,\biggl[p_2^2,\biggl[\frac{\alpha}{r}\biggr]_\epsilon\biggr]\biggr]
-\frac{\alpha}{8\,m^2}\,\biggl\{
\bigl[p_1^i,V\bigr]\,\biggl[\frac{r^i\,r^j-3\,\delta^{ij}\,r^2}{r}\biggr]_\epsilon\,
\bigl[V,p_2^j\bigr]\nonumber \\ &&
+\bigl[p_1^i,V\bigr]\,\biggl[\frac{p_2^2}{2\,m},
\biggl[\frac{r^i\,r^j-3\,\delta^{ij}\,r^2}{r}\biggr]_\epsilon\biggr]\,p_2^j
+p_1^i\,\biggl[\biggl[\frac{r^i\,r^j-3\,\delta^{ij}\,r^2}{r}\biggr]_\epsilon,
\frac{p_1^2}{2\,m}\biggr]\,\bigl[V,p_2^j\bigr]
\nonumber \\ &&
+p_1^i\,\biggl[\frac{p_2^2}{2\,m},\biggl[
\biggl[\frac{r^i\,r^j-3\,\delta^{ij}\,r^2}{r}\biggr]_\epsilon,
\frac{p_1^2}{2\,m}\biggr]\biggr]\, p_2^j\biggr\},
\label{2.51}
\end{eqnarray}
where $[(r^i\,r^j-3\,\delta^{ij}\,r^2)/r]_\epsilon$ is defined in Eq. (\ref{A28}).

$\delta H_8$ is the retardation correction to single transverse
photon exchange, where one vertex is nonrelativistic, Eq. (\ref{2.13})
and the second comes from the 5$^{\rm th}$ term in Eq. (\ref{2.1})
\begin{equation}
-\frac{e}{8\,m^2}\,\sigma^{ij}\bigl\{E^i\,,\,p^j\bigr\}.
\end{equation}
With the help of Eq. (\ref{2.48}) one obtains the following expression for $\delta E_8$
\begin{eqnarray}
\delta E_8 &=& e^2\,\int\frac{d^dk}{(2\,\pi)^d}\,
\biggl(\delta^{ij}-\frac{k^i\,k^j}{k^2}\biggr)\,\frac{i}{16\,m^3}\,
\langle\phi|\,\sigma_1^{ik}\,\bigl\{e^{i\,\vec k\cdot\vec r_1}\, ,\,p_1^k\bigr\}
\nonumber \\ &&\frac{1}{E_0-H_0-k}\,\biggl(p_2^j -\frac{i}{2}\sigma_2^{lj}\,k^l\biggr)
\,e^{-i\,\vec k\cdot\vec r_2}\,|\phi\rangle
+{\rm h.c.} + (1 \leftrightarrow 2).
\end{eqnarray}
In the expansion of $1/(E_0-H_0-k)$ in Eq. (\ref{2.49}) the first term vanishes
because it cancels out with its hermitian conjugate and the second term is a correction of order $m\,\alpha^6$.
After commuting $(H_0-E_0)$ on the left one obtains the effective operator $\delta H_8$
\begin{eqnarray}
\delta H_8 &=&\sum_a \frac{e^2}{8\,m^2}\,\sigma_a^{ij}
\bigl\{{\cal E}_a^i\,,\,{\cal A}_{a}^j\bigr\}
-\frac{i\,e}{16\,m^3}\,\bigl[\sigma_a^{ij}\,\bigl\{p_a^i\,,\,{\cal A}_a^j\bigr\}\,,\,p_a^2\bigr]
\nonumber \\ &=&
\frac{\sigma_1\,\sigma_2}{d}\,\biggl\{
-\frac{Z\,\alpha^2}{8}\biggl[\frac{\vec r_1}{r_1^3} - \frac{\vec
    r_2}{r_2^3}\biggr]_\epsilon\,\biggl[\frac{\vec r}{r^3}\biggr]_\epsilon
+\frac{1}{4}\,\biggl[\frac{\alpha^2}{r^4}\biggr]_\epsilon 
\nonumber \\ &&
+\frac{1}{32}\,\biggl[p_1^2,\biggl[p_1^2,\biggl[\frac{\alpha}{r}\biggr]_\epsilon\biggr]\biggr]
+\frac{1}{32}\,\biggl[p_2^2,\biggl[p_2^2,\biggl[\frac{\alpha}{r}\biggr]_\epsilon\biggr]\biggr]
\biggr\}.
\end{eqnarray}

$H_H$  is the high energy contribution which is given by the forward
three-photon exchange scattering amplitude. It was calculated for 
$m\,\alpha^6$ correction to the parapositronium binding energy in \cite{pos}.
Following \cite{ay}, we define $d$-dimensional spin wave function in analogy to 
this parapositronium work (see Eq. (\ref{3.29})) and take the result with reversed sign 
\begin{equation}
H_H = \biggl(-\frac{1}{\epsilon}+4\,\ln m -\frac{39\,\zeta(3)}{\pi^2}
+\frac{32}{\pi^2}-6\,\ln(2)+\frac{7}{3}\biggr)\,
\frac{\pi\,\alpha^3}{4\,m^2}\,\delta^d(r), \label{2.55}
\end{equation}
where $\zeta$ is the Riemann $\zeta$-function, and we follow the convention that a common factor
$[(4\,\pi)^\epsilon\,\Gamma(1+\epsilon)]^2$ is pulled out from all matrix elements.

$H_R$ is a radiative correction
and its derivation requires a separate treatment. This is based on our former work
for helium \cite{helium1}, and this result has also been obtained in Ref. \cite{ky}.
It is a sum of one- and two-loop contributions
\begin{eqnarray}
H_{R} &=& H_{R1}+H_{R2},
\nonumber \\ 
H_{R1} &=& \frac{\alpha\,(Z\,\alpha)^2}{m^2}\,
\biggl[\frac{427}{96}-2\,\ln(2)\biggr]\,\pi\,[\delta^3(r_1)+\delta^3(r_2)]
\nonumber \\ &&
           +\frac{\alpha^3}{m^2}\,\biggl[
\frac{6\,\zeta(3)}{\pi^2}-\frac{697}{27\,\pi^2}-8\,\ln(2)+\frac{1099}{72}
\biggr]\,\pi\,\delta^3(r), \label{2.56}\\
H_{R2} &=& \frac{\alpha^2\,(Z\,\alpha)}{m^2}\,
\biggl[-\frac{9\,\zeta(3)}{4\,\pi^2}-\frac{2179}{648\,\pi^2}+\frac{3\,\ln(2)}{2}-\frac{10}{27}\biggr]
\,\pi\,[\delta^3(r_1)+\delta^3(r_2)]
\nonumber \\ &&
           +\frac{\alpha^3}{m^2}\,\biggl[
\frac{15\,\zeta(3)}{2\,\pi^2}+\frac{631}{54\,\pi^2}-5\,\ln(2)+\frac{29}{27}
\biggr]\,\pi\,\delta^3(r), \label{2.57}
\end{eqnarray}
At this point we have obtained all contributions of the order of $m\,\alpha^6$.

\section{Elimination of Singularities}
The elimination of singularities will be performed in atomic units,
which in $d$-dimensions become little more complicated.
The nonrelativistic Hamiltonian in natural units is
\begin{equation}
H_0 = \frac{\vec{p_1}^{2}}{2\,m} + \frac{\vec{p_2}^{2}}{2\,m}
-Z\,\alpha\,\frac{C_1}{r_1^{1-2\,\epsilon}}
-Z\,\alpha\,\frac{C_1}{r_2^{1-2\,\epsilon}}
+\alpha\,\frac{C_1}{r_{12}^{1-2\,\epsilon}}.
\end{equation}
Using coordinates in atomic units
\begin{equation}
\vec r \rightarrow (m\,\alpha)^{-1/(1+2\,\epsilon)}\,\vec r, \label{3.2}
\end{equation}
it can be written as
\begin{equation}
H_0 = m^{(1-2\,\epsilon)/(1+2\epsilon)}\,\alpha^{2/(1+2\,\epsilon)}\,
\biggl[\frac{\vec{p_{1}}^{2}}{2} + \frac{\vec{p_{2}}^{2}}{2}
-Z\,\frac{C_1}{r_1^{1-2\,\epsilon}}
-Z\,\frac{C_1}{r_2^{1-2\,\epsilon}}
+\frac{C_1}{r_{12}^{1-2\,\epsilon}}\biggr].
\end{equation}
If one pulls out the factor
$m^{(1-2\,\epsilon)/(1+2\epsilon)}\,\alpha^{2/(1+2\,\epsilon)}$
from $H_0$, than one will obtain the nonrelativistic Hamiltonian in atomic units.
Similarly for $H^{(6)}$, the common factor in atomic units 
\begin{equation}
m^{(1-10\,\epsilon)/(1+2\epsilon)}\,\alpha^{6/(1+2\,\epsilon)}, \label{3.4}
\end{equation}
is pulled out from all the terms, which corresponds to the replacement
$m\rightarrow 1, \alpha \rightarrow 1$. Such a factor will also be pulled out
from $H_H$ in Eq. (\ref{2.55}), which will lead to appearance of logarithmic terms. 

We will explore now the power of dimensional regularization.
All integrals of the form
\begin{equation}
\int d^d k\, k^\alpha \equiv 0,
\end{equation}
vanish identically by definition. We will use this fact in the following.
Consider the matrix element
\begin{equation}
\biggl\langle\phi\biggl|\delta^d(r)\,\biggl[\frac{1}{r}\biggr]_\epsilon
\biggr|\phi\biggr\rangle
= \int \frac{d^d p_1}{(2\,\pi)^d}\,\phi^\star(p_1)\,
\int\frac{d^d p_2}{(2\,\pi)^d}\,\int\frac{d^d p_3}{(2\,\pi)^d}\,
\frac{4\,\pi}{(\vec p_2 - \vec p_3)^2}\,\phi(p_3),
\end{equation} 
and change the variable $\vec p_2 = \vec q_2 + \vec p_3$, then
\begin{equation}
\biggl\langle\phi\biggl|\delta^d(r)\,\biggl[\frac{1}{r}\biggr]_\epsilon
\biggr|\phi\biggr\rangle
= \int \frac{d^d p_1}{(2\,\pi)^d}\,\phi^\star(p_1)\,
\int\frac{d^d q_2}{(2\,\pi)^d}\,\frac{4\,\pi}{q_2^2}\,\int\frac{d^d p_3}{(2\,\pi)^d}\,
\phi(p_3) = 0\,.
\end{equation} 
The integral with $q_2$ vanishes, so this matrix element is equal to $0$,
similarly
\begin{equation}
\biggl\langle\phi\biggl|\delta^d(r_a)\,\biggl[\frac{1}{r_a}\biggr]_\epsilon
\biggr|\phi\biggr\rangle = 0\,.
\end{equation}
Let us introduce momenta $\vec P$ and $\vec p$
\begin{eqnarray}
\vec p_1 &=& \frac{\Vec P}{2}+\vec p,\\
\vec p_2 &=& \frac{\Vec P}{2}-\vec p,
\end{eqnarray}
where $p_a$ is a momentum of the electron $a$
and consider the matrix element
\begin{equation}
\bigl\langle\phi\bigl|\vec p\,\delta^d(r)\,\vec p\bigr|\phi\bigr\rangle
= \int \frac{d^d P}{(2\,\pi)^d}\,
\biggl|\int \frac{d^d p}{(2\,\pi)^d}\, \phi(\vec P,\vec p)\,\vec p\,\biggr|^2 = 0\,.
\end{equation}
This is equal to $0$ because the integrand $\phi(\vec P,\vec p)\,\vec p$ is odd in
$\vec p$ for the singlet states considered here.
These matrix elements and the Schr\"odinger equation 
\begin{equation}
\biggl(\frac{\vec P^2}{4} + \vec p^{\,2} +V\biggr)\phi = E\,\phi,
\end{equation}
are used to derive various identities, for example
\begin{eqnarray}
\langle\nabla^2\delta^d(r)\rangle &=& -\langle\,[\vec p,[\vec p,\delta^d(r)]]\,\rangle
\nonumber \\ &=&
-2\,\langle\delta^d(r)\,p^2\rangle 
\nonumber \\ &=&
-2\,\biggl\langle\delta^3(r)\,\biggl(E+\frac{Z}{r_1} + \frac{Z}{r_2}
-\frac{\vec P^2}{4}\biggr)\biggr\rangle.
\end{eqnarray}
Similarly
\begin{eqnarray}
\biggl[\frac{Z^2}{r_1^4}\biggr]_\epsilon &\equiv& 
\biggl(\nabla_1\biggl[\frac{Z}{r_1}\biggr]_\epsilon\biggr)^2 = 
\vec p_1\,\frac{Z^2}{r_1^2}\,\vec p_1
-2\,\biggl(E+\frac{Z}{r_2}-\frac{1}{r}-\frac{\vec p_2^{\,2}}{2}\biggr)\,
\frac{Z^2}{r_1^2} -2\,\biggl[\frac{Z}{r_1}\biggr]_\epsilon^3,
\label{3.10}\\
\biggl[\frac{1}{r^4}\biggr]_\epsilon &\equiv& 
\biggl(\nabla\biggl[\frac{1}{r}\biggr]_\epsilon\biggr)^2 =
\frac{1}{2}\,\vec p_1\,\frac{1}{r^2}\vec p_1 
+\frac{1}{2}\,\vec p_2\,\frac{1}{r^2}\vec p_2 
-\biggl(E+\frac{Z}{r_1}+\frac{Z}{r_2}\biggr)
\,\frac{1}{r^2} + \biggl[\frac{1}{r}\biggr]_\epsilon^3.
\label{3.11}
\end{eqnarray}
Since the electron-nucleus divergences cancel out algebraically, one does not
need the matrix element of $\langle 1/r_1^3\rangle$. However
$\langle 1/r^3\rangle$ is needed and is calculated as follows
\begin{eqnarray}
\biggl\langle\biggl[\frac{1}{r}\biggr]_\epsilon^3\biggr\rangle &=& 
C_1^3\,\int d^d r\,\phi^2(r)\,r^{-3+6\,\epsilon}
\nonumber \\ &=&
C_1^3\,\phi^2(0)\,\int^\varepsilon d^d r\,r^{-3+6\,\epsilon} +
\int_\varepsilon d^3 r\,\phi^2(r)\,r^{-3}
\nonumber \\ &=&
\biggl\langle\frac{1}{r^3}\biggr\rangle +
\bigl\langle\pi\,\delta^d(r)\bigr\rangle\,
\biggl(\frac{1}{\epsilon}+2\biggr),
\label{3.12}
\end{eqnarray}
where $\langle 1/r^3 \rangle \equiv P(1/r^3)$ is defined in Eq. (\ref{1.5}).
The matrix elements of $1/r^4$ can be obtained from Eq. (\ref{3.11}),
but it can also be calculated directly,
\begin{eqnarray}
\biggl\langle\biggl[\frac{1}{r^4}\biggr]_\epsilon\biggr\rangle &=& 
C_1^2\,\int d^d r\,\phi^2(r)\,[\nabla(r^{-1+2\,\epsilon})]^2
\nonumber \\ &=&
C_1^2\,(-1+2\,\epsilon)^2\phi^2(0)\,\int^\varepsilon d^d r\,r^{-4+4\,\epsilon}\, 
(1-C_2\,r^{1+2\,\epsilon})^2
+\int_\varepsilon d^3 r\,\phi^2(r)\,r^{-4}
\nonumber \\ &=&
\biggl\langle\frac{1}{r^4}\biggr\rangle +
\bigl\langle\pi\,\delta^d(r)\bigr\rangle\,
\biggl(\frac{1}{\epsilon}-4\biggr),
\label{3.13}
\end{eqnarray}
where we assume that $1/\varepsilon$ and $\ln\varepsilon+\gamma$ are dropped.
Similarly
\begin{equation}
\biggl[\frac{1}{2\,r}\,\biggl(\delta^{ij}+\frac{r^i\,r^j}{r^2}\biggr)\biggr]_\epsilon\,
\nabla^i\,\nabla^j\,\biggl[\frac{1}{r}\biggr]_\epsilon = 
\frac{1}{r^4} +\pi\,\delta^d(r)\,\biggl(\frac{1}{\epsilon}-5\biggr).
\end{equation}
From Eqs. (\ref{3.12}) and (\ref{3.13}) one obtains the identity
\begin{equation}
\biggl\langle\frac{1}{r^4}\biggr\rangle =
\biggl\langle\frac{1}{r^3}\biggr\rangle 
+ \vec p_1\,\frac{1}{2\,r^2}\,\vec p_1
+ \vec p_2\,\frac{1}{2\,r^2}\,\vec p_2
-\biggl(E+\frac{Z}{r_1}+\frac{Z}{r_2}\biggr)\,\frac{1}{r^2} + 6\,\pi\,\delta^{3}(r).
\label{3.14}
\end{equation}
Other identities which will be used are
\begin{eqnarray}
4\,\pi\,\delta_\perp^{ij}\,p^i\,p^j &=& -\pi\nabla^2\,\delta^d(r) 
-\frac{Z}{4}\,\biggl(\frac{\vec r_1}{r_1^3}-\frac{\vec r_2}{r_2^3}\biggr)\cdot\frac{\vec r}{r^3}
+\frac{1}{2}\,\biggl[\frac{1}{r^4}\biggr]_\epsilon, \\
4\,\pi\,\delta_\perp^{ij}\,P^i\,P^j &=& 
P^i\,P^j\,\frac{(3\,r^i\,r^j - \delta^{ij}\,r^2)}{r^5}
+\frac{8\,\pi}{3}\,\delta^{3}(r)\,P^2,\\
\biggl[p_1^2,\biggl[p_1^2,\biggl[\frac{1}{r}\biggr]_\epsilon\biggr]\biggr] &=& 
Z\,\biggl(\frac{\vec r_1}{r_1^3}-\frac{\vec r_2}{r_2^3}\biggr)\cdot\frac{\vec r}{r^3}
-2\,\biggl[\frac{1}{r^4}\biggr]_\epsilon 
+\frac{4}{3}\,\pi\,\delta^d(r)\,P^2 
\nonumber \\ &&
- P^i\,P^j\,\frac{(3\,r^i\,r^j-\delta^{ij}\,r^2)}{r^5},\\
\biggl[p_2^2,\biggl[p_1^2,\biggl[\frac{1}{r}\biggr]_\epsilon\biggr]\biggr] &=& 
Z\,\biggl(\frac{\vec r_1}{r_1^3}-\frac{\vec r_2}{r_2^3}\biggr)\cdot\frac{\vec r}{r^3}
-2\,\biggl[\frac{1}{r^4}\biggr]_\epsilon 
-\frac{4}{3}\,\pi\,\delta^d(r)\,P^2 
\nonumber \\ &&
+ P^i\,P^j\,\frac{(3\,r^i\,r^j-\delta^{ij}\,r^2)}{r^5},\\
p_1^2\,\biggl[\frac{1}{r}\biggr]_\epsilon\,p_2^2 &=& 
\frac{1}{r}\,\biggl(E+\frac{Z}{r_1} + \frac{Z}{r_2}\biggr)^2
-\frac{2}{r^2}\,\biggl(E+\frac{Z}{r_1} + \frac{Z}{r_2}\biggr) 
+\biggl[\frac{1}{r}\biggr]^3_\epsilon 
\nonumber \\ &&
-\vec P\cdot\vec p\,\frac{1}{r}\,\vec p\cdot\vec P,\\
\vec p_1\cdot\vec p_2\,\biggl[\frac{1}{r}\biggr]_\epsilon\,\vec p_1\cdot\vec p_2
&=&\frac{1}{r}\,\biggl(\frac{P^2}{2}-\frac{Z}{r_1}-\frac{Z}{r_2}+\frac{1}{r}-E\biggr)^2
   + \pi\,\delta^d(r)\,\biggl(\frac{1}{\epsilon}+2\biggr), \\
\vec P\cdot\vec p\,\frac{1}{r}\,\vec p\cdot\vec P &=& 
-\vec p_1\times \vec p_2\,\frac{1}{r}\,\vec p_1\times \vec p_2 - 
\frac{P^4}{4\,r} +
\frac{P^2}{r}\,\biggl(E+\frac{Z}{r_1}+\frac{Z}{r_2}-\frac{1}{r}\biggr)
\nonumber \\ &&
-2\,\pi\,\delta^{3}(r)\,P^2,
\label{3.21}
\end{eqnarray}
We are now ready to eliminate divergences from matrix elements of $\delta H_i$
operators. In the following we make replacement $\sigma_1\,\sigma_2\rightarrow -d(d-1)$.
To show this, we consider matrix elements of spin operators with the helium singlet
wave function in $d$- dimensions.
All terms with a single spin operator vanish. The spin-spin operators of the
form $\sigma_1^{ij}\,\sigma_2^{ik}\,(\delta^{jk}/d -r^j\,r^k/r^2)$ vanish
because of angular integration. Only operators of the form
$\sigma_1^{ij}\,\sigma_2^{ij}\,Q $ fail to vanish, and we use a definition which was implicitly
assumed in Ref. \cite{ky}. Namely, we postulate the existence of the
charge conjugation operator $C$, such that
\begin{eqnarray}
C^{-1} &=& C^T, \\
C\,\vec \sigma_2^T\,C^{-1} &=& -\vec \sigma_1,
\end{eqnarray}
and singlet states $\psi_S$ are defined by
\begin{equation}
\langle\psi_S| Q_1 \otimes Q_2|\psi_S\rangle = \frac{1}{2}\,
{\rm Tr}\,\bigl[Q_1\,C\,Q_2^T\,C^{-1}\bigr]. \label{3.29}
\end{equation}
Using this definition one obtains
\begin{equation}
\langle\psi_S| Q\,\sigma_1^{ij} \sigma_2^{ij}|\psi_S\rangle = -d(d-1)\,
\langle\psi_S| Q |\psi_S\rangle,
\label{B16}
\end{equation}
for an arbitrary operator $Q$. 

All $\delta E_i$ corrections are now transformed
as follows. The first term, $\delta E_0 = \langle\delta H_0\rangle$ becomes
\begin{eqnarray}
\delta E_0 &=& \frac{1}{16}\,\bigl\langle p_1^6 + p_2^6\bigr\rangle 
 = \frac{1}{16}\,\bigl\langle (p_1^2 + p_2^2)^3 -3\,p_1^2\,p_2^2\,(p_1^2+p_2^2)\bigr\rangle
\\ &=&
\frac{1}{16}\,\biggl\langle
4\,\bigl[(\nabla_1 V)^2 + (\nabla_2 V)^2\bigr] + 8\,(E-V)^3 - 6\,p_1^2\,(E-V)\,p_2^2 
+ 3\,\biggl[p_2^2,\biggl[p_1^2,\biggl[\frac{1}{r}\biggr]_\epsilon\biggr]\biggr]\biggr\rangle.
\nonumber
\end{eqnarray}
All singular operators in $E_0$ can be handled by
Eqs. (\ref{3.10}) - (\ref{3.21}), and all the
singularities are identified in the form of $\phi^2(0)/\epsilon$.
Next the terms $\delta E_1-\delta E_8$ are transformed in a similar way
\begin{eqnarray}
\delta E_1 &=& \frac{1}{128}\,\biggl\{
-4\bigl[(\nabla_1 V)^2 + (\nabla_2 V)^2\bigr]
-2\,\biggl[p_2^2,\biggl[p_1^2,\biggl[\frac{1}{r}\biggr]_\epsilon\biggr]\biggr]\biggr\}
\nonumber \\ &&
-\frac{3}{32}\,\biggl\{
2\,\biggl(E+\frac{Z-1}{r_2}\biggr)\,4\,\pi\,Z\,\delta^3(r_1) + 
2\,\biggl(E+\frac{Z-1}{r_1}\biggr)\,4\,\pi\,Z\,\delta^3(r_2)
\nonumber \\ &&
-2\,\biggl(E+\frac{Z}{r_1}+\frac{Z}{r_2}\biggr)\,4\,\pi\,Z\,\delta^3(r)
-p_2^2\,4\,\pi\,Z\,\delta^3(r_1) - p_1^2\,4\,\pi\,Z\,\delta^3(r_2)\biggr\},
\nonumber \\
\delta E_2 &=& -\frac{\pi}{16}\,\nabla^2\,\delta^d(r) 
-\frac{\pi}{16}\,\delta_\perp^{ij}\,P^i\,P^j
+\frac{\pi}{4}\,\delta_\perp^{ij}\,p^i\,p^j, \\ 
\nonumber \\
\delta E_3 &=& -\frac{1}{16}\,\biggl[p_2^2,\biggl[p_1^2,\biggl[\frac{1}{r}\biggr]_\epsilon\biggr]\biggr],\\
\nonumber \\
\delta E_4 &=& p_1^i\,(E-V)\frac{1}{2\,r}\,\biggl(\delta^{ij}+\frac{r^i\,r^j}{r^2}\biggr)\,p_2^j
-\frac{1}{2}\,\biggl(E+\frac{Z}{r_1}+\frac{Z}{r_2}\biggr)\,4\,\pi\delta^3(r)
\nonumber \\ &&
-\frac{1}{2}\,\biggl[\frac{1}{2\,r}\,\biggl(\delta^{ij}+\frac{r^i\,r^j}{r^2}\biggr)\biggr]_\epsilon\,
\nabla^i\,\nabla^j\,\biggl[\frac{1}{r}\biggr]_\epsilon, \\
\nonumber \\
\delta E_5 &=& \frac{Z}{4}\,
\biggl(\frac{\vec r_1}{r_1^3}-\frac{\vec r_2}{r_2^3}\biggr)\cdot\frac{\vec r}{r^3} 
-\frac{(d-1)}{4}\,\biggl[\frac{1}{r^4}\biggr]_\epsilon,\\ \nonumber \\
\delta E_6 &=& \frac{1}{8}\,p_1^i\,\frac{1}{r^2}\,\biggl(\delta^{ij}+3\frac{r^i\,r^j}{r^2}\biggr)\,p_1^j
+\frac{1}{8}\,p_2^i\,\frac{1}{r^2}\,\biggl(\delta^{ij}+3\frac{r^i\,r^j}{r^2}\biggr)\,p_2^j
+\frac{d-1}{4}\,\biggl[\frac{1}{r^4}\biggr]_\epsilon,\\
\nonumber \\
\delta E_8 &=& \frac{Z}{4}\,\biggl(\frac{\vec r_1}{r_1^3} - \frac{\vec r_2}{r_2^3}\biggr)
\cdot\frac{\vec r}{r^3}
-\frac{(d-1)}{4}\,\biggl[\frac{1}{r^4}\biggr]_\epsilon 
\nonumber \\ &&
-\frac{(d-1)}{32}\,\biggl[p_1^2,\biggl[p_1^2,\biggl[\frac{1}{r}\biggr]_\epsilon\biggr]\biggr]
-\frac{(d-1)}{32}\,\biggl[p_2^2,\biggl[p_2^2,\biggl[\frac{1}{r}\biggr]_\epsilon\biggr]\biggr].
\end{eqnarray}
The calculation of the $E_7$ contribution is lengthier, and we split it
into 4 parts, corresponding to Eq. (\ref{2.51})
\begin{equation}
\delta E_7 = \delta E_{7A} + \delta E_{7B} + \delta E_{7C} + \delta E_{7D}. 
\end{equation}
Each part contains singular operators which are handled similarly to that in
Eqs. (\ref{3.10}) - (\ref{3.21})
\begin{eqnarray}
\delta E_{7A} &=&
-\frac{(d-1)}{16}\,\biggl[p_2^2,\biggl[p_1^2,\biggl[\frac{1}{r}\biggr]_\epsilon\biggr]\biggr],
\\ \nonumber \\
\delta E_{7B} &=& -\biggl(Z\,\frac{r_1^i}{r_1^3}-\frac{r^i}{r^3}\biggr)\,
\biggl(Z\,\frac{r_2^j}{r_2^3}+\frac{r^i}{r^3}\biggr)\,
\frac{(r^i\,r^j-3\,\delta^{ij}\,r^2)}{8\,r} 
-\frac{\pi}{4}\,\delta^d(r)\,\biggl(\frac{1}{\epsilon}-5\biggr),
\\ \nonumber \\
\delta E_{7C} &=& -\frac{Z}{8}\,p_2^k\,\frac{r_1^i}{r_1^3}\,
\biggl(\delta^{jk}\,\frac{r^i}{r} - \delta^{ik}\,\frac{r^j}{r} 
-\delta^{ij}\,\frac{r^k}{r} - \frac{r^i\,r^j\,r^k}{r^3}\biggr)\,p_2^j
+\frac{1}{8}\,p_2^j\,\frac{(\delta^{jk}\,r^2 - 3 r^j\,r^k)}{r^4}\,p_2^k
\nonumber \\ &&
-\frac{Z}{8}\,p_1^k\,\frac{r_2^i}{r_2^3}\,
\biggl(\delta^{jk}\,\frac{r^i}{r} - \delta^{ik}\,\frac{r^j}{r} 
-\delta^{ij}\,\frac{r^k}{r} - \frac{r^i\,r^j\,r^k}{r^3}\biggr)\,p_1^j
+\frac{1}{8}\,p_1^j\,\frac{(\delta^{jk}\,r^2 - 3 r^j\,r^k)}{r^4}\,p_1^k
\nonumber \\ &&
+\frac{1}{4\,r^4} + \frac{\pi}{4}\,\delta^d(r)\,\biggl(\frac{1}{\epsilon}-7\biggr),
\\\nonumber \\
\delta E_{7D} &=& \frac{1}{8}\,\vec p_1\times\vec p_2\,\frac{1}{r}\,\vec p_1\times\vec p_2 
-\frac{1}{8}\,\vec p_1\cdot\vec p_2\,\biggl[\frac{1}{r}\biggr]_\epsilon\,\vec p_1\cdot\vec p_2
\nonumber \\ &&
+\frac{1}{8}\,p_1^k\,p_2^l\,\biggl(- \frac{\delta^{jl}\,r^i\,r^k}{r^3}
- \frac{\delta^{ik}\,r^j\,r^l}{r^3}
+ 3\,\frac{r^i\,r^j\,r^k\,r^l}{r^5} \biggr)\,p_1^i\,p_2^j. 
\end{eqnarray}
At this point we have completed the elimination of singularities from
the effective Hamiltonian. It remains to consider, however, the second order matrix element $E_A$   
\begin{equation}
E_A = \biggl\langle H_A\,\frac{1}{(E_0-H_0)'}\,H_A\biggr\rangle,
\end{equation}
which requires subtractions of  $1/\epsilon$ singularities.
For this we use the transformation
\begin{eqnarray}
H_A &=& H'_A +  \bigl\{H_0-E_0,Q\bigr\}, \\
Q &=& -\frac{1}{4}\biggl[\frac{Z}{r_1}+\frac{Z}{r_2}\biggr]_\epsilon 
+ \frac{(d-1)}{4}\,\biggl[\frac{1}{r}\biggr]_\epsilon,
\end{eqnarray}
so that 
\begin{eqnarray}
E_A &=& E'_A  + E''_A,\\
E'_A &=& \biggl\langle H'_A\,\frac{1}{(E_0-H_0)'}\,H'_A\biggr\rangle, \label{3.39}\\
E''_A &=& \bigl\langle Q\,(E_0-H_0)\,Q\bigr\rangle
+ 2\,\langle H_A\rangle\,\langle Q\rangle
-\bigl\langle \bigl\{H_A\,,\,Q\bigr\}\bigr\rangle = X_1+X_2+X_3.
\end{eqnarray}
$E'_A$ is finite in the limit $\epsilon \rightarrow 0$, and
\begin{eqnarray}
H'_A |\phi\rangle &=& \biggl\{
-\frac{1}{2}\,(E_0-V)^2
-p_1^i\,\frac{1}{2\,r}\,\biggl(\delta^{ij}+\frac{r^i\,r^j}{r^2}\biggr)\,p_2^j
+\frac{1}{4}\,\vec \nabla_1^2 \, \vec \nabla_2^2
\nonumber \\ &&
-\frac{Z}{4}\,\frac{\vec r_1}{r_1^3}\cdot\vec \nabla_1
-\frac{Z}{4}\,\frac{\vec r_1}{r_1^3}\cdot\vec \nabla_1
\biggr\}|\phi\rangle,
\end{eqnarray}
where $\vec \nabla_1^2 \, \vec \nabla_2^2$ is understood as a differentiation
of $\phi$ on the right hand side as a function (omitting $\delta^3(r)$). 
What remains is the calculation of $X_i$ terms. The first two are simple
\begin{eqnarray}
X_1 &=& \frac{1}{32}\,\biggl[\frac{Z^2}{r_1^4} +
\frac{Z^2}{r_2^4}\biggr]_\epsilon
+\frac{(d-1)^2}{16}\,\biggl[\frac{1}{r^4}\biggr]_\epsilon
-\frac{Z}{8}\,\biggl(\frac{\vec r_1}{r_1^3} - \frac{\vec r_2}{r_2^3}\biggr)
\cdot\frac{\vec r}{r^3},\\
X_2 &=& 2\,E^{(4)}\,\biggl\langle
-\frac{1}{4}\biggl(\frac{Z}{r_1}+\frac{Z}{r_2}\biggr)
+ \frac{1}{2\,r}\biggr\rangle.
\end{eqnarray}
To calculate $X_3$ we split it again into four parts correspondingly
\begin{eqnarray}
X_3 &=& -2\langle\phi|\biggl\{-\frac{p_1^4}{8} - \frac{p_2^4}{8} 
+\frac{\pi\,Z}{2}\,\bigl[\delta^d(r_1)+\delta^d(r_2)\bigr]\nonumber \\ &&
-\frac{1}{2}\,p_1^i\,
\biggl[\frac{\delta^{ij}}{r}+\frac{r^i\,r^j}{r^3}
\biggr]_\epsilon\,p_2^j
+\pi\,(d-2)\,\delta^d(r)\biggr\}
\nonumber \\ &&
\times\biggl\{
-\frac{1}{4}\biggl[\frac{Z}{r_1}+\frac{Z}{r_2}\biggr]_\epsilon 
+ \frac{(d-1)}{4}\,\biggl[\frac{1}{r}\biggr]_\epsilon\biggr\}\,|\phi\rangle
\nonumber \\ &=& X_{3A} + X_{3B} + X_{3C}+ X_{3D},
\end{eqnarray}
and calculate each part separately
\begin{eqnarray}
X_{3A} &=& \frac{1}{4}\,\bigl\langle(p_1^4+p_2^4)\,Q\,\bigr\rangle
\nonumber \\ &=&
\frac{1}{4}\,\biggl\langle
(p_1^2+p_2^2)\,Q\,(p_1^2+p_2^2) 
+\frac{1}{2}\,[p_1^2+p_2^2,[p_1^2+p_2^2,Q]]
-2 p_1^2\,Q\,p_2^2 
-[p_1^2,[p_2^2,Q]]\biggr\rangle
\nonumber \\ &=&
\frac{1}{4}\,\biggl(E
+\biggl[\frac{Z}{r_1}+\frac{Z}{r_2}\biggr]_\epsilon
-\biggl[\frac{1}{r}\biggr]_\epsilon\biggr)^2\,\biggl(
(d-1)\,\biggl[\frac{1}{r}\biggr]_\epsilon
-\biggl[\frac{Z}{r_1}+\frac{Z}{r_2}\biggr]_\epsilon\biggr)
\nonumber \\ &&
-\frac{1}{8}\,\biggl(
\biggl[\frac{Z^2}{r_1^4}+ \frac{Z^2}{r_2^4}\biggr]_\epsilon
+2\,(d-1)\,\biggl[\frac{1}{r^4}\biggr]_\epsilon 
-3\,Z\,\biggl(\frac{\vec r_1}{r_1^3}
 -\frac{\vec r_2}{r_2^3}\biggr)\cdot\frac{\vec r}{r^3}\biggr)
\nonumber \\ &&
+\frac{1}{8}\,p_1^2\,\biggl(\frac{Z}{r_1}+\frac{Z}{r_2}\biggr)\,p_2^2
-\frac{(d-1)}{8}\,p_1^2\,\biggl[\frac{1}{r}\biggr]_\epsilon\,p_2^2
-\frac{(d-1)}{16}\,\biggl[p_2^2,\biggl[p_1^2,\biggl[\frac{1}{r}\biggr]_\epsilon\biggr]\biggr],
\\ \nonumber \\
X_{3B} &=& Z\,(Z-2)\,\frac{\pi}{4}\,\delta^3(r_1)\,\frac{1}{r_2}
+Z\,(Z-2)\,\frac{\pi}{4}\,\delta^3(r_2)\,\frac{1}{r_1},\\ \nonumber \\
X_{3C} &=& -\frac{1}{4}\,p_1^i\,\biggl(
\frac{Z}{r_1}+\frac{Z}{r_2}-\frac{2}{r}\biggr)\,\frac{1}{r}\,
\biggl(\delta^{ij}+\frac{r^i\,r^j}{r^3}\biggr)\,p_2^j + \frac{(d-1)}{4}\,
\biggl[\frac{1}{2\,r}\,\biggl(\delta^{ij}+\frac{r^i\,r^j}{r^2}\biggr)\biggr]_\epsilon\,
\nabla^i\,\nabla^j\,\biggl[\frac{1}{r}\biggr]_\epsilon,
\nonumber \\ \\
X_{3D} &=& \frac{\pi}{2}\,\delta^3(r)\,\biggl(\frac{Z}{r_1} + \frac{Z}{r_2}\biggr).
\end{eqnarray}

The last term to be considered is $H_H$ in Eq. (\ref{2.55}). One transforms it
to atomic units by the replacement in Eq. (\ref{3.2}) and division by factor
in Eq. (\ref{3.4}). As a result one obtains
\begin{eqnarray}
H_H &=& \biggl(-\frac{1}{\epsilon}-4\,\ln \alpha -\frac{39\,\zeta(3)}{\pi^2}
+\frac{32}{\pi^2}-6\,\ln(2)+\frac{7}{3}\biggr)\, 
\frac{\pi}{4}\,\delta^d(r)\nonumber \\
&=& H'_H - \biggl(\frac{1}{\epsilon}+4\,\ln \alpha\biggr)
\,\frac{\pi}{4}\,\delta^d(r). \label{3.58}
\end{eqnarray}
At this point we have separated out all singularities.
They always have the form of $\phi^2(0)/\epsilon$, and finally
cancel between themselves. The sum of all  terms, which is the main result of
this work, is
\begin{eqnarray}
E^{(6)} &=& -\frac{E_0^3}{2} 
+\biggl[\biggl(-E_0 + \frac{3}{2}\,p_2^2 
+\frac{1-2\,Z}{r_2}\biggr)\,\frac{Z\,\pi}{4}\,\delta^{3}(r_1)
+(1\leftrightarrow 2)\biggr]
\nonumber \\ &&
+\biggl( 1-\frac{Z}{r_1} - \frac{Z}{r_2} +\frac{P^2}{3}\biggr)\,\frac{\pi}{2}\,\delta^{3}(r)
+\frac{E_0^2+2\,E^{(4)}}{4\,r} -\frac{E_0}{2\,r^2} +\frac{1}{4\,r^3}
\nonumber \\ &&
-\frac{E_0}{2\,r}\,\bigg(\frac{Z}{r_1}+\frac{Z}{r_2}\biggr)
+\frac{E_0}{4}\,\biggl(\frac{Z}{r_1}+\frac{Z}{r_2}\biggr)^2
-\frac{1}{4\,r^2}\,\biggl(\frac{Z}{r_1}+\frac{Z}{r_2}\biggr)
-\frac{1}{4\,r}\,\biggl(\frac{Z}{r_1}+\frac{Z}{r_2}\biggr)^2
\nonumber \\ &&
+\frac{Z^2}{2\,r_1\,r_2}\,\biggl(E_0+\frac{Z}{r_1}+\frac{Z}{r_2}-\frac{1}{r}\biggr)
+\frac{Z}{32}\,\biggl(\frac{\vec r_1}{r_1^3} - \frac{\vec r_2}{r_2^3}\biggr)\cdot\frac{\vec r}{r^3}
+\frac{Z}{4}\,\biggl(\frac{\vec r_1}{r_1^3} - \frac{\vec r_2}{r_2^3}\biggr)\cdot\frac{\vec r}{r^2}
\nonumber \\ &&
-\frac{Z^2}{8}\,\frac{r_1^i}{r_1^3}\,\frac{(r^i r^j - 3\,\delta^{ij}\,r^2)}{r}\,\frac{r_2^j}{r_2^3}
+\biggl[\frac{Z^2}{8}\,\frac{1}{r_1^2}\,\vec p_2^{\,2} + \frac{Z^2}{8}\,\vec p_1\,\frac{1}{r_1^2}\,\vec p_1
+\frac{1}{2}\,\vec p_1\,\frac{1}{r^2}\,\vec p_1
+(1\leftrightarrow 2)\biggr]
\nonumber \\ &&
+\frac{1}{4}\,p_1^i\,\biggl(\frac{Z}{r_1}+\frac{Z}{r_2}\biggr)\,
\frac{(r^i\,r^j + \delta^{ij}\, r^2)}{r^3}\, p_2^j
-\frac{1}{32}\, P^i\,\frac{(3\,r^i\,r^j - \delta^{ij} r^2)}{r^5}\,P^j
\nonumber \\ &&
-\biggl[\frac{Z}{8}\,p_2^k\,\frac{r_1^i}{r_1^3}\,\biggl(\frac{\delta^{jk}\,r^i}{r} 
- \frac{\delta^{ik}\, r^j}{r} - \frac{\delta^{ij}\, r^k}{r} 
- \frac{r^i\, r^j\, r^k}{r^3}\biggr)\,p_2^j
+(1\leftrightarrow 2)\biggr]
\nonumber \\ &&
-\frac{E_0}{8}\,p_1^2\,p_2^2-\frac{1}{4}\,p_1^2\,\biggl(\frac{Z}{r_1}
+\frac{Z}{r_2}\biggr)\,p_2^2
+\frac{1}{4}\,\vec p_1\times\vec p_2\,\frac{1}{r}\,\vec p_1\times\vec p_2
\nonumber \\ &&
+\frac{1}{8}\,p_1^k\,p_2^l\,\biggl(-\frac{\delta^{jl}\,r^i\,r^k}{r^3} -
           \frac{\delta^{ik}\,r^j\,r^l}{r^3} +
           3\,\frac{r^i\,r^j\,r^k\,r^l}{r^5} \biggr)\, p_1^i\,p_2^j
\nonumber \\ &&
+ E'_H + E'_A + E_C + E_{R1} + E_{R2} -\ln(\alpha)\,\pi\,\delta^d(r), \label{3.49}
\end{eqnarray}
where $E'_H = \langle H'_H\rangle$ from Eq. (\ref{3.58}), $E'_A$ is defined in
Eq. (\ref{3.39}), $E_C$ in Eq. (\ref{4.12}), $E_{R1}$ and $E_{R2}$ in
Eqs. (\ref{2.56}), (\ref{2.57}) correspondingly.
In addition to various identities in Eqs. (\ref{3.14})-(\ref{3.21}), 
we used two further equations
\begin{eqnarray}
\biggl\langle \frac{Z}{r_1}+\frac{Z}{r_2}\biggr\rangle &=& 
\biggl\langle\frac{1}{r}\biggr\rangle-2 E_0,\\
\biggl\langle p_1^i\,\frac{(r^i\,r^i +
  \delta^{ij}\,r^2)}{r^3}\,p_2^j\biggr\rangle &=&
-2\,E^{(4)} -\biggl(E_0+\frac{Z}{r_1}+\frac{Z}{r_2}-\frac{1}{r}\biggr)^2
+\frac{p_1^2\,p_2^2}{2}
\nonumber \\ &&
+\pi\,Z\,\bigl[\delta^{3}(r_1) + \delta^{3}(r_2)\bigr]
+2\,\pi\,\delta^{3}(r),
\end{eqnarray}
to simplify final expression. The logarithmic term in Eq. (\ref{3.49})
agrees with that obtained in Ref. \cite{lna}.
The sum of ``soft'' operators (Eq. (\ref{3.49}) without the last line)  
after $(1 \leftrightarrow 2)$ simplification becomes
\begin{eqnarray}
E_Q &=& -\frac{E_0^3}{2} - \frac{E_0\,Z}{8}\,\,Q_{1} + \frac{1}{8}\,Q_{2} 
            -\frac{Z\,(2\,Z - 1)}{8}\,Q_{3} + \frac{3\,Z}{16}\,Q_{4} 
            -\frac{Z}{4}\,Q_{5} + \frac{1}{24}\,Q_{6} 
\nonumber \\ &&
            +\frac{E_0^2 + 2\,E^{(4)}}{4}\,Q_{7} - \frac{E_0}{2}\,Q_{8} + \frac{1}{4}\,Q_{9} 
            +\frac{E_0\,Z^2}{2}\,Q_{11} + E_0\,Z^2\,Q_{12} - E_0\,Z\,Q_{13} 
\nonumber \\ &&
            -Z^2\,Q_{14} + Z^3\,Q_{15} - \frac{Z^2}{2}\,Q_{16} - \frac{Z}{2}\,Q_{17} 
            +\frac{Z}{16}\,Q_{18} + \frac{Z}{2}\,Q_{19} - \frac{Z^2}{8}\,Q_{20} + \frac{Z^2}{4}\,Q_{21}
\nonumber \\ &&
            +\frac{Z^2}{4}\,Q_{21} + \frac{Z^2}{4}\,Q_{22} + Q_{23} + \frac{Z}{2}\,Q_{24} 
            -\frac{1}{32}\,Q_{25} - \frac{Z}{4}\,Q_{26} - \frac{E_0}{8}\,Q_{27} -\frac{Z}{2}\,Q_{28} 
\nonumber \\ &&
+ \frac{1}{4}\,Q_{29} + \frac{1}{8}\,Q_{30},
\end{eqnarray}
where $Q_i$ are defined in Table I.

\section{Numerical calculations of matrix elements}
The helium wave function is expanded in a basis set of
exponential functions in the form of \cite{kor}
\begin{equation}
\phi(r_1,r_2,r) = \sum_{i=1}^{\cal N} v_i
[e^{-\alpha_i r_1-\beta_i r_2-\gamma_i r} + (r_1 \leftrightarrow r_2)],
\label{4.1}
\end{equation}
where $\alpha_i$, $\beta_i$ and $\gamma_i$ are generated randomly
with conditions:
\begin{eqnarray}
A_1<\alpha_i<A_2,\;\; \beta_i+\gamma_i>\varepsilon, \\
B_1<\beta_i<B_2,\;\; \alpha_i+\gamma_i>\varepsilon, \\
C_1<\gamma_i<C_2,\;\; \alpha_i+\beta_i>\varepsilon.
\end{eqnarray}
In order to obtain a highly precise wave function following
Korobov \cite{kor}, we use double set of the form (\ref{4.1}). 
Each parameters $A_i,B_i,C_i,\varepsilon$ are determined  by the energy
minimization, with the condition that $\varepsilon >0$, which is 
necessary for the normalizability of the wave function.
The linear coefficients $v_i$ in Eq. (\ref{4.1}) form
a vector $v$, which is a solution of the generalized 
eigenvalue problem
\begin{equation}
H v = E N v,
\end{equation}
where $H$ is a matrix of the Hamiltonian in this basis set,
$N$ is a normalization (overlap) matrix, and $E$ an eigenvalue, 
the energy of the state corresponding to $v$. For the solution
of the eigenvalue problem with ${\cal N} = 100, 300, 600, 900, 1200, 1500$ 
we use LU decomposition in quad and sextuple precision. 
As a result we obtain the following nonrelativistic energies in au
\begin{eqnarray}
E_0(1^1S_0) &=& -2.903~724~377~034~119~592(6), \label{4.6}\\
E_0(2^1S_0) &=& -2.145~974~046~054~417~311(50). \label{4.7}
\end{eqnarray}
These values agree with the more accurate result of 
Korobov \cite{kor} and of Drake in \cite{Drake_h}.
The calculation of matrix elements of nonrelativistic Hamiltonian
 can be performed with the use of one formula:
\begin{equation}
\frac{1}{16\,\pi^2}\,\int d^3 r_1\,\int d^3r_2\,
\frac{e^{-\alpha r_1-\beta r_2-\gamma r}}{r_1\,r_2\,r} =
\frac{1}{(\alpha+\beta)(\beta+\gamma)(\gamma+\alpha)}.
\end{equation}
The result with any additional powers of $r_i$ in the numerator can be 
obtained by differentiation with respect to the corresponding parameter 
$\alpha$, $\beta$ or $\gamma$. The matrix elements of relativistic corrections
involve inverse powers of $r_1, r_2, r$. 
These, can be obtained by integration with respect to corresponding parameter.
This leads to the appearance of logarithmic and dilogarithmic
functions, for example
\begin{eqnarray}
\frac{1}{16\,\pi^2}\,\int d^3 r_1\,\int d^3r_2\,
\frac{e^{-\alpha r_1-\beta r_2-\gamma r}}{r_1\,r_2\,r^2} &=&
\frac{1}{(\beta + \alpha)\,(\alpha + \beta)}\,
\ln\left(\frac{\beta + \gamma}{\alpha + \gamma}\right)\,,
\\
\frac{1}{16\,\pi^2}\,\int d^3 r_1\,\int d^3r_2\,
\frac{e^{-\alpha r_1-\beta r_2-\gamma r}}{r_1^2\,r_2\,r^2} &=&
\frac{1}{2\,\beta}\,\biggl[
\frac{\pi^2}{6} + \frac{1}{2}\,\ln^2\left(\frac{\alpha + \beta}
{\beta + \gamma}\right)  \nonumber \\ && + 
{\rm Li}_2\left(1 - \frac{\alpha + \gamma}{\alpha + \beta}\right) + 
{\rm Li}_2\left(1 - \frac{\alpha + \gamma}{\beta + \gamma}\right)\biggr]\,.
\end{eqnarray}
All matrix elements involved in the $m\,\alpha^6$ correction, see Table I, can be expressed
in terms of rational, logarithmic and dilogarithmic functions, as above.
The high quality of the wave function allows us to obtain precise matrix elements
of $H^{(6)}$ operators and the numerical results are presented in Table I.
\begin{table}[htb]
\renewcommand{\arraystretch}{0.95}
\caption{\label{tabledi} Expectation values of operators entering $H^{(6)}$ for
  1S state, $\vec r= \vec r_1 - \vec r_2 $}
\label{TBL1}
\begin{tabular}{rl@{\hspace{0.1cm}}.}
\hline
\hline
$Q_{1}=$ &$ 4\,\pi\,\delta^{3}(r_1)         $& 22x.750~526 \\
$Q_{2}=$ &$ 4\,\pi\,\delta^{3}(r)           $&  1x.336~375 \\
$Q_{3}=$ &$ 4\,\pi\,\delta^{3}(r_1)/r_2     $& 33x.440~565 \\
$Q_{4}=$ &$ 4\,\pi\,\delta^{3}(r_1)\,p_2^2  $& 49x.160~046 \\
$Q_{5}=$ &$ 4\,\pi\,\delta^{3}(r)/r_1       $&  5x.019~714 \\
$Q_{6}=$ &$ 4\,\pi\,\delta^{3}(r)\,P^2      $& 18x.859~765 \\
$Q_{7}=$ &$ 1/r                             $&  0x.945~818 \\
$Q_{8}=$ &$ 1/r^2                           $&  1x.464~771 \\
$Q_{9}=$ &$ 1/r^3                           $&  0x.989~274 \\
$Q_{10}=$&$ 1/r^4                           $& -3x.336~383 \\
$Q_{11}=$&$ 1/r_1^2                         $&  6x.017~409 \\
$Q_{12}=$&$ 1/(r_1\,r_2)                    $&  2x.708~655 \\
$Q_{13}=$&$ 1/(r_1\,r)                      $&  1x.920~944 \\
$Q_{14}=$&$ 1/(r_1\,r_2\,r)                 $&  4x.167~175 \\
$Q_{15}=$&$ 1/(r_1^2\,r_2)                  $&  9x.172~094 \\
$Q_{16}=$&$ 1/(r_1^2\,r)                    $&  8x.003~454 \\
$Q_{17}=$&$ 1/(r_1\,r^2)                    $&  3x.788~791 \\
$Q_{18}=$&$ (\vec r_1\cdot \vec r)/(r_1^3\,r^3) 
                                            $&  3x.270~472 \\ 
$Q_{19}=$&$ (\vec r_1\cdot \vec r)/(r_1^3\,r^2) 
                                            $&  1x.827~027 \\
$Q_{20}=$&$ r_1^i\,r_2^j\,(r^i r^j - 3\,\delta^{ij}\,r^2)/(r_1^3\,r_2^3\,r)  
                                            $&  0x.784~425 \\
$Q_{21}=$&$ p_2^2/r_1^2                     $& 14x.111~960 \\
$Q_{22}=$&$ \vec p_1\, /r_1^2\, \vec p_1    $& 21x.833~598 \\
$Q_{23}=$&$ \vec p_1\, /r^2\, \vec p_1      $&  4x.571~652 \\
$Q_{24}=$&$ p_1^i\,(r^i\,r^j + \delta^{ij}\, r^2)/(r_1\,r^3)\, p_2^j            
                                            $& 0x.811~933  \\
$Q_{25}=$&$ P^i\,(3\,r^i\,r^j - \delta^{ij} r^2)/r^5\,P^j                           
                                            $& -3x.765~488 \\
$Q_{26}=$&$  p_2^k\, r_1^i\,/r_1^3\,(\delta^{jk}\, r^i/r - \delta^{ik}\, r^j/r - 
           \delta^{ij}\, r^k/r - r^i\, r^j\, r^k/r^3)\,p_2^j
                                            $& -0x.266~894 \\
$Q_{27}=$&$ p_1^2\,p_2^2                    $&  7x.133~710 \\
$Q_{28}=$&$ p_1^2\,/r_1\,p_2^2              $& 37x.010~642 \\
$Q_{29}=$&$ \vec p_1\times\vec p_2\,/r\,\vec p_1\times\vec p_2           
                                            $&  4x.004~703 \\
$Q_{30}=$&$ p_1^k\,p_2^l\,(-\delta^{jl}\,r^i\,r^k/r^3 -
           \delta^{ik}\,r^j\,r^l/r^3 + 3\,r^i\,r^j\,r^k\,r^l/r^5)\, p_1^i\,p_2^j   
                                            $& -1x.591~864 \\
\hline
\hline
\end{tabular}
\end{table}
Some of these matrix elements have already been calculated in \cite{drake_op}
and results in Table I are in agreement with them.
 
The calculation of second order corrections $E'_A$ and $E_C$ is more complicated.
The spin algebra in the second order matrix element $E_C$ 
is simplified with the help of
\begin{equation}
|^1S_0\rangle\,\langle ^1S_0| = |S_0\rangle\,\langle S_0|\,\biggl(1-\frac{\vec s^{\,2}}{2}\biggr),
\end{equation}
where $\langle \vec r_1, \vec r_2|S_0\rangle$ is the wave function without the
spin, and $\vec s = (\vec \sigma_1 + \vec\sigma_2)/2$, so that
\begin{eqnarray}
E_C &=& \biggl\langle S_0\biggl|\vec C\,\frac{1}{E_0-H_0}\,\vec C\biggr|
S_0\biggr\rangle, \label{4.12}\\ 
\vec C &=&\frac{Z}{4}\biggl(\frac{\vec r_1}{r_1^3}\times\vec p_1 
- \frac{\vec r_2}{r_2^3}\times\vec p_2\biggr)
+\frac{1}{4}\,\frac{\vec r}{r^3}\times(\vec p_1+\vec p_2). 
\label{4.13}
\end{eqnarray}
The inversion of the operator $E_0-H_0$ in this expression
is performed in a basis set of even parity functions with $l=1$ of the form
\begin{equation}
\vec \phi(r_1,r_2,r) = \sum_i v_i\,\vec r_1 \times \vec r_2\,
[e^{-\alpha_i r_1-\beta_i r_2-\gamma_i r} + (r_1 \leftrightarrow r_2)]\,,
\end{equation}
The values of parameters $A_i$, $B_i$ and $C_i$ corresponding to $\vec \phi$
are obtained by minimization of $E_C$, and results of these calculations 
are presented in Table II.
\begin{table}[!htb]
\renewcommand{\arraystretch}{1.0}
\caption{Contributions to $E^{(6)}$ for 1S and 2S states of the helium
  atom. $E_{LG}$ is the logarithmic correction, last term in Eq. (\ref{3.49}).}
\label{TBL3}
\begin{ruledtabular}
\begin{tabular}{c@{\extracolsep{\fill}}w{3.11}w{3.11}w{3.11}}
$m\,\alpha^6$         & \centt{He($1^1$S)}&\centt{He($2^1$S)}&
\centt{$\Delta E$} \\
\hline
$E_Q$                 &  15.465\,431      &  12.310\,132     & -3.155\,299    \\
$E'_H$                &  -0.278\,403      &  -0.022\,641     &  0.255\,762    \\
$E'_A$                & -18.495\,345(50)  & -16.280\,186(10) &  2.215\,159(50)\\
$E_C$                 &  -0.392\,621      &  -0.033\,790     &  0.358\,831    \\
\hline
Subtotal              &  -3.700\,937(50)  &  -4.026\,485(10) & -0.325\,547(50)\\ 
$E_{R1}$              & 141.924\,288      & 100.971\,873     &-40.952\,415    \\
$E_{R2}$              &   1.144\,012      &   0.890\,559     & -0.253\,453    \\
$E_{LG}$              &   1.643\,823      &   0.133\,682     & -1.510\,141    \\
\hline
Total                 & 141.011\,185(50)  &  97.969\,630(10) &-43.041\,555(50)\\ 
$-E_D({\rm He}^+)$    &   4.000\,000      &   4.000\,000     &  0.000\,000    \\
$-E_{R1}({\rm He}^+)$ & -97.971\,914      & -97.971\,914     &  0.000\,000    \\
$-E_{R2}({\rm He}^+)$ &  -0.873\,699      &  -0.873\,699     &  0.000\,000    \\
\hline
$E^{(6)}({\rm He}) - E^{(6)}({\rm He}^+)$      
                      &  46.165\,572(50)  &  3.124\,017(10)  &-43.041\,555(50)\\ 
\end{tabular}
\end{ruledtabular}
\end{table}
The calculation of $E'_A$ is similar to that of $E_C$, but additionally
requires a subtraction of the reference state from the implicit sum over states.
We obtain this by orthogonalization of $H'_A |\phi\rangle$ with respect to 
eigenstate with closest to $0$ eigenvalue of $H-E$. This eigenvalue is not
exactly equal to $0$, because we use a basis set with different parameters,
which are obtained by minimization of $E'_A$. Results are presented in Table
II. Surprisingly, the total nonlogarithmic exchange contribution after subtraction of He$^+$
value is very small, namely  $0.299063$ for 1S state and $-0.026485$ for 2S
state. This contribution is much smaller than the dominating one-loop contribution $E_{R1}$
which is $43.952\,374$ and $2.999\,959$ correspondingly.
It is similar for triplet states $2^3S_1$ and $2^3 P_1$ of helium,
and means that higher order corrections can be well approximated by 
the one-loop self-energy contribution.

\section{Summary}
We have derived the complete order $m\,\alpha^6$ contribution
to energy levels of singlet states of helium.
It is expressed as the expectation value of the operators  
in Eq. (\ref{3.49}). A similar, but not identical set of operators
have been obtained previously by Yelkhovsky, Eq. (97) in \cite{ay},
and the results obtained here are in slight disagreement (see Appendix C for details).
The matrix elements of operators entering Eq. (\ref{3.49}) for the ground state
of helium atom are presented in Table I, and a few of them are 
in disagreement with the results presented in Ref. \cite{ky}.
Because of these discrepancies, calculations presented here should be verified
before definite conclusions can be made.

In this work we performed numerical calculations for the ground $1^1S_0$ and excited 
$2^1 S_0$ states, and the results are presented in Table II.
While the calculation of the $E_Q$ operators was quite complicated,
their contribution to $E^{(6)}$ is relatively small. The dominating
contribution comes from the one-loop electron self-energy $E_{R1}$, and is given by
Dirac delta functions, see Eq. (\ref{2.56}). 

The summary of all known contributions to $1^1S_0 - 2^1 S_0$ transition is
presented in Table III.
\begin{table}[!htb]
\renewcommand{\arraystretch}{1.3}
\caption{Contributions to 1S and 2S ionization energies of the helium atom in MHz. 
         Physical constants from \cite{nist}, Ry$=10973731.568525(73)$ m$^{-1}$, 
         $\alpha  = 1/137.03599911(46)$, $\not\!\lambda_e = 386.1592678(26)$ fm,
         $m_\alpha/m_e = 7294.2995363(32)$, $r_\alpha = 1.673$ fm, $c =
         299792458$. The uncertainty for $E^{(7)}$ is due to its approximate
         calculation and is roughly estimated to be about half of $E^{(7)}$}
\label{TBL2}
\begin{ruledtabular}
\begin{tabular}{c@{\extracolsep{\fill}}w{11.8}w{10.8}w{10.8}}
             & \centt{$\nu(1^1S)$} & \centt{$\nu(2^1S)$} & \centt{$\Delta \nu(2^1S- 1^1S)$} \\
\hline
$E^{(2)}$            & -5\,945\,262\,288.61   & -960\,322\,874.90     & 4\,984\,939\,413.71 \\
$E^{(4)}$            &           16\,800.32   &       -11\,974.80     &         -28\,775.12 \\
$E^{(5)}$            &           40\,495.81   &         2\,755.14     &         -37\,740.68 \\
$E^{(6)}$            &               861.24   &             58.28     &             -802.96 \\
$E^{(7)}$            &               -72.(36) &             -4.(2)    &              68.(34)\\
$E_{\rm FS}$         &                29.58   &              1.99     &              -27.59 \\ \hline
Theory               & -5\,945\,204\,174.(36) & -960\,332\,038.(2)& 4\,984\,872\,136.(34)  \\
V.K. and A.Y. \cite{ky}&-5\,945\,204\,223.(42)&                       & \\
Drake \cite{Drake_h} & -5\,945\,204\,223.(91) & -960\,332\,041.(25)   & 4\,984\,872\,182.(91)\\ \hline
Exp.  \cite{bergeson}& -5\,945\,204\,356.(48) & -960\,332\,041.01(15) & 4\,984\,872\,315.(48)\\
Exp. \cite{eikema}   & -5\,945\,204\,238.(45) &                       &                      \\
\end{tabular}
\end{ruledtabular}
\end{table}
The nonrelativistic energy here, $E^{(2)}$ is the sum of $\mu E_0$ 
from Eqs. (\ref{4.6}) and (\ref{4.7}) with $\mu$ being the reduced mass, 
and mass polarization corrections from Ref. \cite{Drake_h}.
Relativistic contribution $E^{(4)}$ is taken from Ref. \cite{martin} and includes
nuclear recoil. Leading QED contribution $E^{(5)}$ also includes nuclear recoil \cite{helrec},
and we use Bethe logarithms as obtained by Drake in Ref. \cite{drake_bethe}. Our value
for $E^{(5)}$ is greater by about $12$ MHz  from  that of Korobov and
Yelkhovsky in Ref. \cite{ky}, which is $40\,483.98(5)$ MHz, and we do not
understand a reason of this discrepancy.
We have not been able to find in the literature the result of Drake for $E^{(5)}$ 
as well as separate results for higher order terms. However, 
the total result of Drake, see Table III, is in agreement with that of Ref. \cite{ky}.
$E^{(6)}$ is obtained here and our result is greater by about 25 MHz
from the result in Ref. \cite{ky}, which is $834.9(2)$. 
The source of this deviation is explained in Appendix C. $E^{(7)}$ includes
all electron-nucleus terms of order $m\,\alpha^7$ which are known
from the hydrogen Lamb shift \cite{eides} (one-, two- and three-loop
contributions) and are extended to helium in the standard way.
Our value is larger by $12$ MHz from the result of Ref. \cite{ky} $-84(42)$, 
because we include all $\alpha^7$ terms, not only the leading
$\ln^2(\alpha)$. The current theoretical uncertainty
comes mainly from the approximate treatment of these higher order terms.
The exact calculation of  $E^{(7)}$ is at present very difficult, due to high
complexity in the derivation of $H^{(7)}$ and thus limits the accuracy of theoretical
predictions. Our final theoretical predictions are in moderate agreement with the
measurement of Eikema {\em et al} \cite{eikema}, and disagree significantly with 
the measurement by Bergeson {\em et al} in \cite{bergeson}.

Having the exact formula for $m\,\alpha^6$ corrections for singlet and as well as
triplet states \cite{triplet}, it is possible now to improve theoretical predictions
for higher excited states of helium, as well as light helium-like ions.
The extension of this approach to three- and more electron atoms or molecules
is possible, but not all technical problems in calculating matrix elements in
explicitly correlated basis set have been resolved yet.

\section*{Acknowledgments}
I wish to acknowledge valuable comments from V. Korobov.

\appendix
\section{Dimensionally regularized QED of bound states}
The dimension of space is assumed to be $d=3-2\,\epsilon$. 
The  surface area of $d$-dimensional unit sphere 
can be obtained by considering the following integral
\begin{equation}
{\rm I} = \int d^d k\,e^{-\vec k^2}.
\end{equation}
In Cartesian coordinates it is a product of $d$ one-dimensional integrals
\begin{equation}
{\rm I} = \biggl[\int dk \, e^{-k^2}\biggr]^d = \pi^{d/2}, \label{A2}
\end{equation}
while in spherical coordinates it is
\begin{equation}
{\rm I} = \int d \Omega_d\,\int_0^\infty dk\,k^{d-1}\,e^{-k^2} = 
\Omega_d\,\frac{1}{2}\,\Gamma(d/2). \label{A3}
\end{equation}
From comparison of Eq. (\ref{A2}) and (\ref{A3}) one obtains
\begin{equation}
\Omega_d = \frac{2\,\pi^{d/2}}{\Gamma(d/2)}\,.
\end{equation}
The $d$-dimensional Laplacian is $\nabla^2 = \partial^i\,\partial^i$.
For spherically symmetric functions $f$, $g$
\begin{eqnarray}
\int d^d r\,\nabla^2(f)\, g  &=& -\int d^d r\,\nabla(f)\cdot\nabla(g) = 
-\Omega_d\,\int dr\,r^{d-1}\,\partial_r f\,\partial_r g \nonumber \\ &=& 
\Omega_d\,\int dr\,\partial_r(r^{d-1}\,\partial_r f)\,g =
\int d^d r\,r^{1-d}\,\partial_r(r^{d-1}\,\partial_r f)\,g\,,
\end{eqnarray}
Laplacian takes the form
\begin{equation}
\nabla^2= r^{1-d}\partial_r\,r^{d-1}\partial_r.
\end{equation}
The photon propagator, and thus Coulomb interaction preserves its form in the momentum
representation, while in the coordinate representation it is 
\begin{equation}
{\cal V}(r) = \int\frac{d^d k}{(2\,\pi)^d}\,\frac{4\,\pi}{k^2}\,e^{i\,\vec k\cdot\vec r}
= \pi^{\epsilon-1/2}\,\Gamma(1/2-\epsilon)\,r^{2\,\epsilon-1} \equiv\frac{C_1}{r^{1-2\,\epsilon}}.
\end{equation} 
The alternative derivation of ${\cal V}(r)$ which omits the calculation of Fourier transform is 
the following. Consider the equation
\begin{equation}
\nabla^2 {\cal V}(r) = -4\,\pi\delta^d(r).
\end{equation}
If one assumes that ${\cal V}(r)$ is of the form ${\cal V}(r) = C\,r^\gamma$, then for
$r\neq 0$
\begin{equation}
r^{1-d}\partial_r\,r^{d-1}\partial_r (C\,r^\gamma) = 0,
\end{equation}
and therefore $\gamma = 2-d = 2\,\epsilon-1$. The coefficient $C$ is obtained
by considering the integral with the trial function $f$
\begin{eqnarray}
4\,\pi\,f(0) &=& -\int d^d r\, \nabla^2({\cal V})\,f = \int d^d r\, \nabla({\cal V})\cdot
\nabla(f) \nonumber \\ &=&
 \lim_{\varepsilon\rightarrow 0}\int_\varepsilon dr\,r^{d-1}\,
\Omega_d\,\partial_r({\cal V})\,\partial_r(f) \nonumber \\ &=&
 \lim_{\varepsilon\rightarrow 0}\biggl[
r^{d-1}\,\Omega_d\,\partial_r({\cal V})\,f\biggr|_{r=\varepsilon}
-\int_\varepsilon dr\,\Omega_d\,
\partial_r(r^{d-1}\,\partial_r({\cal V}))\,f\biggr] \nonumber \\ &=&
 \lim_{\varepsilon\rightarrow 0}\,
\Omega_d\,\varepsilon^{d-1}\,\partial_\varepsilon(C\,\varepsilon^{2-d})\,f(\varepsilon)
 \nonumber \\ &=& \Omega_d\,(2-d)\,C\,f(0),
\end{eqnarray}
therefore 
\begin{equation}
C \equiv C_1 = \frac{4\,\pi}{(d-2)\,\Omega_d} = \pi^{\epsilon-1/2}\,\Gamma(1/2-\epsilon).
\end{equation}
We are now ready to consider quantum mechanics in $d$-dimensions
The nonrelativistic Hamiltonian of the hydrogen-like systems is
\begin{equation}
H_0 = \frac{\vec{p}^{\;2}}{2}
-Z\,\frac{C_1}{r^{1-2\,\varepsilon}}\,,
\end{equation}
and of helium-like systems
\begin{equation}
H_0 = \frac{\vec{p_1}^{2}}{2} + \frac{\vec{p_2}^{2}}{2}
-Z\,\frac{C_1}{r_1^{1-2\,\epsilon}}
-Z\,\frac{C_1}{r_2^{1-2\,\epsilon}}
+\frac{C_1}{r_{12}^{1-2\,\epsilon}}.
\end{equation}
The solution of stationary Schr\"odinger equation 
$H_0\,\phi = E_0\,\phi$ we denote by $\phi$, and will never refer to its 
explicit (and unknown) form in $d$-dimensions. Instead, we will use only
the generalized cusp condition to eliminate various singularities
from matrix elements with relativistic operators. Namely,
we expect, that for small $r\equiv r_{12}$
\begin{equation}
\phi(r)\approx \phi(0)\,(1-C\,r^\gamma),
\end{equation}
with some coefficient $C$ and $\gamma$ to be obtained from the
two-electron Schr\"odinger equation around $r=0$
\begin{equation}
[-\nabla^2 + {\cal V}(r)]\phi(0)\,(1-C\,r^\gamma) \approx E\,\phi(0)\,(1-C\,r^\gamma).
\end{equation}
From the  cancellation of small $r$ singularities of the left hand side
of above equation, one obtains
\begin{eqnarray}
\gamma &=& 1+2\,\epsilon,\\
C \equiv C_2 &=& \frac{1}{4}\,\pi^{\epsilon-1/2}\,\Gamma(-1/2-\epsilon). 
\label{A17}
\end{eqnarray}
Therefore, the two-electron wave function around $r_{12}=0$ behaves as
\begin{eqnarray}
\phi(\vec r_1,\vec r_2) \approx \phi(r_{12}=0)\,
\bigl(1-C_2\,r_{12}^{1+2\,\epsilon}\bigr).
\end{eqnarray}

Apart from the Coulomb potential ${\cal V}(r)$ in the coordinate space, 
we need also other functions, which appear in the calculations of 
relativistic operators, namely
\begin{eqnarray}
{\cal V}_2(r) &=& \int \frac{d^d k}{(2\,\pi)^d}\,\frac{4\,\pi}{k^4}\,
           e^{i\,\vec k\cdot\vec r},\\
{\cal V}_3(r) &=& \int \frac{d^d k}{(2\,\pi)^d}\,\frac{4\,\pi}{k^6}\,
           e^{i\,\vec k\cdot\vec r}.
\end{eqnarray}
They can be obtained from the differential equations
\begin{eqnarray}
-\nabla^2 {\cal V}_2(r) &=& {\cal V}(r),\\
-\nabla^2 {\cal V}_3(r) &=& {\cal V}_2(r),\\
\end{eqnarray}
with the results
\begin{eqnarray}
{\cal V}_2(r) &=& C_2\,r^{1+2\,\epsilon},\\
{\cal V}_2(r) &=& C_3\,r^{3+2\,\epsilon},\\
\end{eqnarray}
with $C_2$ defined in Eq. (\ref{A17}) and
\begin{equation}
C_3 = \frac{1}{32}\,\pi^{\epsilon-1/2}\,\Gamma(-3/2-\epsilon).
\end{equation}
Using ${\cal V}_i$ we calculate various integrals involving photon propagator
in the Coulomb gauge, namely
\begin{eqnarray}
&& \int \frac{d^d k}{(2\,\pi)^d}\,\frac{4\,\pi}{k^4}\,
\biggl(\delta^{ij}-\frac{k^i\,k^j}{k^2}\biggr)\,e^{i\,\vec k\cdot\vec r}
=\delta^{ij} {\cal V}_2 + \partial^i\partial^j\,{\cal V}_3\nonumber \\
&=&\pi^{\epsilon-1/2}\,r^{-1+2\,\epsilon}\,
\biggl[\frac{3}{16}\,\delta^{ij}\,\Gamma(-1/2-\epsilon)\,r^2
+\frac{1}{8}\,\Gamma(1/2-\epsilon)\,r^i\,r^j\biggr]\nonumber\\
& \equiv & \biggl[\frac{1}{8\,r}(r^i\,r^j-3\,\delta^{ij}\,r^2)\biggr]_\epsilon,
\label{A28}
\end{eqnarray}
and
\begin{eqnarray}
&& \int \frac{d^d k}{(2\,\pi)^d}\,\frac{4\,\pi}{k^2}\,
\biggl(\delta^{ij}-\frac{k^i\,k^j}{k^2}\biggr)\,e^{i\,\vec k\cdot\vec r}
=\delta^{ij} {\cal V} + \partial^i\partial^j\,{\cal V}_2\nonumber \\
&=&\pi^{\epsilon-1/2}\,r^{-3+2\,\epsilon}\,
\biggl[\frac{1}{2}\,\delta^{ij}\,\Gamma(1/2-\epsilon)\,r^2
+\Gamma(3/2-\epsilon)\,r^i\,r^j\biggr]\nonumber\\
& \equiv & \biggl[\frac{1}{2\,r^3}(\delta^{ij}\,r^2 + r^i\,r^j)\biggr]_\epsilon,
\end{eqnarray}
and
\begin{eqnarray}
&& \int \frac{d^d k}{(2\,\pi)^d}\,4\,\pi\,
\biggl(\delta^{ij}-\frac{k^i\,k^j}{k^2}\biggr)\,e^{i\,\vec k\cdot\vec r}
=\frac{(d-1)}{d} \delta^{ij} 4\,\pi\,\delta^d(r) +
\biggl(\partial^i\partial^j-\frac{\delta^{ij}}{d}\,\partial^2\biggr)\,{\cal V}
\nonumber \\
&=&\frac{(d-1)}{d} \delta^{ij} 4\,\pi\,\delta^d(r) +
\pi^{\epsilon-1/2}\,r^{-5+2\,\epsilon}\,
\biggl[-2\,\delta^{ij}\,\Gamma(3/2-\epsilon)\,r^2
+4\,\Gamma(5/2-\epsilon)\,r^i\,r^j\biggr]\nonumber\\
&=& \biggl[\frac{2}{3}\,\delta^{ij}\,4\,\pi\,\delta^3(r) + 
\frac{1}{r^5}\,(3\,r^i\,r^j-\delta^{ij}\,r^2)\biggr]_\epsilon \equiv \delta_\perp^{ij}.
\label{A30}
\end{eqnarray}

\section{Foldy-Wouthuysen transformation in $d$-dimensions}
The Foldy-Wouthuysen (FW) transformation \cite{itzykson, fw} is the
nonrelativistic expansion of the Dirac Hamiltonian in an external
electromagnetic field. Here we extend this transformation to 
the arbitrary dimension $d$ of space.
The Dirac Hamiltonian in the external electromagnetic field is
\begin{equation}
H = \vec \alpha \cdot \vec \pi +\beta\,m + e\,A^0\,,
\end{equation}
where $\vec \pi = \vec p-e\,\vec A$,
\begin{equation}
\alpha^i = \left(
\begin{array}{cc}
0&\sigma^i\\
\sigma^i&0
\end{array}\right),\;\;
\beta = 
\left(
\begin{array}{cc}
I&0\\
0&-I
\end{array}\right),
\end{equation}
and
\begin{equation}
\{\sigma^i,\sigma^j\} = 2\,\delta^{ij}\,I.
\end{equation}
The FW transformation $S$ \cite{itzykson} leads to a new Hamiltonian
\begin{equation}
H_{\rm FW} = e^{i\,S}\,(H-i\,\partial_t)\,e^{-i\,S}\,, 
\end{equation}
which decouples the upper and lower components of the Dirac wave
function up to a specified order in the $1/m$ expansion. Here 
we calculate FW Hamiltonian up to terms which contribute
to $m\,\alpha^6$ correction to the energy.  We use a convenient
form of Foldy-Wouthuysen operator $S$, which can be written as 
\begin{eqnarray}
S &=&-\frac{i}{2\,m}\,\biggl\{ \beta\,\vec\alpha\cdot\vec\pi -
\frac{1}{3\,m^2}\,\beta\,(\vec\alpha\cdot\vec\pi)^3
+\frac{1}{2\,m}\,[\vec\alpha\cdot\vec\pi\, ,\,e\,A^0-i\,\partial_t]
+\frac{\beta}{5\,m^4}\,(\vec\alpha\cdot\vec\pi)^5 \nonumber \\ && -
     \frac{\beta\,e}{4\,m^2}\,\vec\alpha\cdot\dot{\vec E}+
     \frac{i\,e}{24\,m^3}\,[\vec\alpha\cdot\vec\pi,
            [\vec\alpha\cdot\vec\pi,\vec\alpha\cdot\vec E]]
     -\frac{i\,e}{3\,m^3}\,\bigl\{(\vec\alpha\cdot\vec\pi)^2\,,\,
            \vec\alpha\cdot\vec E\bigr\}\biggr\}\,.
\end{eqnarray}
The F.W. Hamiltonian is expanded in a power series in $S$
\begin{equation}
H_{\rm FW} = \sum_{j=0}^6 {\cal H}^{(j)}+\ldots \label{09}
\end{equation}
where 
\begin{eqnarray}
{\cal H}^{(0)} &=& H, \nonumber \\
{\cal H}^{(1)} &=& [i\,S\,,{\cal H}^{(0)}-i\,\partial_t],\nonumber \\
{\cal H}^{(j)} &=& \frac{1}{j}\,[i\,S\,,{\cal H}^{(j-1)}]\; {\mbox{\rm for $j>1$}},
\label{10}
\end{eqnarray}
and higher order terms in this expansion, denoted by dots, are
neglected. The calculations of subsequent commutators
is rather tedious but the result simple
\begin{eqnarray}
H_{\rm FW} &=& e\,A^0 + \frac{(\vec\sigma\cdot\vec\pi)^2}{2\,m} - 
\frac{(\vec\sigma\cdot\vec\pi)^4}{8\,m^3}
+\frac{(\vec\sigma\cdot\vec\pi)^6}{16\,m^5}
-\frac{i\,e}{8\,m^2}\,[\vec\sigma\cdot\vec\pi,\vec\sigma\cdot\vec E]
-\frac{e}{16\,m^3}\,\bigl\{\vec\pi\,,\,\partial_t{\vec E}\bigr\}\nonumber \\ && 
-\frac{i\,e}{128\,m^4}\,[\vec\sigma\cdot\vec\pi,[\vec\sigma\cdot\vec\pi,
[\vec\sigma\cdot\vec\pi,\vec\sigma\cdot{\vec E}]]]
+\frac{i\,e}{16\,m^4}\,\Bigl\{(\vec\sigma\cdot\vec\pi)^2\,,\,
[\vec\sigma\cdot\vec\pi,\vec\sigma\cdot\vec E]\Bigr\}\,.
\end{eqnarray}
There is some arbitrariness in the operator $S$, what means
that $H_{\rm FW}$ is not unique. The standard approach \cite{itzykson}, which relies
on subsequent use of FW-transformations differs from this one in $d=3$,
by the transformation $S$ with some additional even operator.

Our aim here is to obtain the Hamiltonian for further calculations
of $m\,\alpha^6$ contribution to energy levels of an arbitrary light atom.
For this one can neglect the vector potential $\vec A$ in all the terms
having $m^4$ and $m^5$ in the denominator. Moreover, less obviously,
one can neglect the term with  $\vec\sigma\cdot\vec A\,\vec\sigma\cdot\dot{\vec E}$ 
and the $\vec B^2$ term. It is because they are of second order in
electromagnetic fields which additionally contain derivatives, 
and thus contribute only at higher orders. 
After these simplifications, $H_{\rm FW}$ takes the form
\begin{eqnarray}
 H_{\rm FW} &=& e\,A^0 + \frac{\pi^2}{2\,m}-\frac{e}{4\,m}\,\sigma^{ij}\,B^{ij} - 
\frac{\pi^4}{8\,m^3}-\frac{e}{8\,m^2}\Bigl(\vec\nabla\cdot\vec E +
\sigma^{ij}\,\bigl\{E^i\,,\,\pi^j\bigr\}\Bigr)
\nonumber \\ &&
+\frac{e}{16\,m^3}\bigl\{\sigma^{ij} B^{ij}\,,\,p^2\bigr\}  
-\frac{e}{16\,m^3}\,\bigl\{\vec p\,,\, \partial_t{\vec E}\bigr\}
+\frac{3\,e}{32\,m^4}\,\bigl\{\sigma^{ij}\,E^i\,p^j\,,\,p^2\bigr\}
\nonumber \\ &&
+\frac{e}{128\,m^4}\,[p^2,[p^2,A^0]]
-\frac{3\,e}{64\,m^4}\,\bigl\{p^2\,,\,\nabla^2 A^0 \bigr\} +\frac{p^6}{16\,m^5},
\end{eqnarray}
where
\begin{eqnarray}
\sigma^{ij} &=& \frac{1}{2\,i}\,[\sigma^i\,,\,\sigma^j],\\
B^{ij} &=& \partial^i\,A^j - \partial^j\,A^i,\\
E^i &=& -\nabla^i A^0 -\partial_t A^i.
\end{eqnarray}

\section{Comparison to former works}
A similar, but not identical set of operators in Eq. (\ref{3.49})
has been obtained by Yelkhovsky, Eq. (97) in \cite{ay}. 
When his operators are transformed by using the following three equations 
\begin{eqnarray}
\biggl\langle\frac{i}{r^3}\,
\biggl(\frac{\vec r}{r}\cdot\vec p-\frac{1}{2}\biggr)\biggr\rangle
 &=&
\frac{1}{4}\,\biggl\langle
\vec p_1\,\frac{1}{r^2}\,\vec p_1 +\vec p_2\,\frac{1}{r^2}\,\vec p_2
-\frac{2}{r^2}\,\biggl(E+\frac{Z}{r_1}+\frac{Z}{r_2}\biggr)+ 4\,\,\pi\,\delta^3(r)
\biggr\rangle, \\
\biggl\langle\frac{i}{r_1^3}\biggl(\frac{\vec r_1}{r_1}\cdot\vec p_1+Z\biggr)\biggr\rangle
&=&
\frac{1}{2}\,\biggl\langle \vec p_1\,\frac{1}{r_1^2}\,\vec p_1 
-\frac{2}{r_1^2}\,\biggl(E+\frac{Z}{r_1}-\frac{1}{r}-\frac{p_1^2}{2}\biggr)
-4\,Z\,\pi\,\delta^3(r_1)\biggr\rangle, 
\end{eqnarray}
\begin{eqnarray}
\frac{1}{8}\,p_1^k\,p_1^i\,\biggl(
-\frac{\delta^{jl}\,r^i\,r^k}{r^3}
-\frac{\delta^{ik}\,r^j\,r^l}{r^3}
+3\,\frac{r^i\,r^j\,r^k\,r^l}{r^5}\biggr)\,p_2^l\,p_2^j &=&
\nonumber \\&& \hspace*{-40ex} 
\frac{1}{8}\,p_1^k\,p_2^l\,\biggl(
-\frac{\delta^{jl}\,r^i\,r^k}{r^3}
-\frac{\delta^{ik}\,r^j\,r^l}{r^3}
+3\,\frac{r^i\,r^j\,r^k\,r^l}{r^5}\biggr)\,p_1^i\,p_2^j
\nonumber \\ && \hspace*{-40ex} 
-\frac{1}{8}\,P^i\,\frac{3\,r^i\,r^i-\delta^{ij}\,r^2}{r^5}\,P^j
+\frac{\pi}{4}\,\delta^3(r)
-\frac{1}{12}\,\pi\,\delta^3(r)\,P^2,
\end{eqnarray}
then almost all operators agree with one exception.
The difference between operators in Eq. (97) of Ref. \cite{ay}
and that of ours Eq. (\ref{3.49}) is
\begin{equation}
\Delta = \frac{1}{32}\,\pi\,\delta^3(r)\,P^2. \label{C4}
\end{equation} 
For this reason we checked the calculation in
Ref. \cite{ay}. The derivation of initial operators was very
similar to our former work in \cite{helium1}. However,
the electron-electron Coulomb interaction, according to
Ref. \cite{ky} involves the term
\begin{equation}
\frac{7\,\pi\,\alpha}{32\,m^4}\,\bigl\{p_1^2+p_2^2,\delta(\vec r)\bigr\},
\label{C5}
\end{equation}
while our calculations in \cite{helium1}  give
\begin{equation}
\frac{6\,\pi\,\alpha}{32\,m^4}\,\bigl\{p_1^2+p_2^2,\delta(\vec r)\bigr\}
-\frac{\alpha}{16\,m^4}\,\nabla^2\,\pi\delta(\vec r).
\label{C6}
\end{equation} 
The difference between Eq. (\ref{C5}) and Eq. (\ref{C6}) for singlet S-states
is equal to $\Delta$ from Eq. (\ref{C4}).
and this term should be subtracted from Eq. (97) of Ref. \cite{ay}. 
Although in this work we use a different formalism
we obtain the same result (after using Schrodinger equation) as in Ref. \cite{helium1}.
namely the sum of our terms $V_C^{eN} + V_C^{ee} + \delta_{C}^{-}V$ 
differs from  $\delta H_1 + \delta H_2 + \delta H_3$ by exactly the same term $\Delta$.

Considering numerical matrix elements, we found that results
presented in Ref. \cite{ky} are not reliable.
Most of them are accurate to only three digits, for example
the matrix element corresponding to $Q_{28}$,
according to KY is $36.983$, while our result is $37.010\,642$. 
Some of them are accurate only to the first digit, for example
for $Q_{29}$ KY gives $4\times 1.078 = 4.312$ and our result is $4.004\,703$.
Some of matrix elements in Ref. \cite{ky} contain misprints in their presentation,
namely they should include additional $1/2$ on the left hand side, to agree with
numerical values and to be consistent with Eq. (97) of Ref. \cite{ay}.
Most importantly, some matrix elements from \cite{ky}
are in error, for example
\begin{equation}
\biggl\langle\frac{3\,i}{2\,r^3}\,
\biggl(\frac{\vec r}{r}\cdot\vec p-\frac{1}{2}\biggr)\biggr\rangle,
\end{equation}
By using integration by parts we transform this matrix element to the form
\begin{equation}
\biggl\langle\frac{3\,i}{2\,r^3}\,
\biggl(\frac{\vec r}{r}\cdot\vec p-\frac{1}{2}\biggr)\biggr\rangle
 =
\frac{3}{4}\,\biggl\langle
\frac{1}{r^4}-\frac{1}{r^3}-4\,\pi\,\delta(r)\biggr\rangle
 = -4.246\,525\,,
\label{5.9}
\end{equation}
which is in disagreement with the result of \cite{ky}, namely  $-0.958$.
Alternatively, we use the identity in Eq. (\ref{3.14})  
and obtain the same numerical value as in Eq. (\ref{5.9}).
Another example is $Q_{26}$, for which result of \cite{ky} is $2\times4.749$
and our result from Table I is $-0.266\,894$.
In conclusion, the numerical matrix elements of Ref. \cite{ky} should be verified.

\end{document}